\documentclass{aa}
\usepackage{graphicx}
\usepackage{latexsym}
\usepackage{color}
\usepackage{ulem}
\usepackage{linenoaa}

\usepackage[utf8]{inputenc}
\usepackage[english]{babel}

\usepackage[dvipsnames]{xcolor}
\usepackage{amssymb}
\usepackage{longtable}    
\usepackage{lscape}
\usepackage{url}
\usepackage{natbib}
\bibpunct{(}{)}{;}{a}{}{,} 
\newcommand{\lamost}{{LAMOST}}

\newcommand{\kepler}{{\it Kepler}}
\newcommand{\ktwo}{{\it K2}}

\newcommand{\teff}{$T_{\rm eff}$}

\newcommand{\logg}{$\log g$}
\newcommand{\vsini}{$v\sin i$}

\newcommand{\feh}{[Fe/H]}

\newcommand{\vrad}{$V_r$}

\newcommand{\rotfit}{{\tt ROTFIT}}

\newcommand{\kms}{km\,s$^{-1}$}
\newcommand{\halpha}{H$\alpha$}
\newcommand{\Whalpha}{$EW^{\rm res}_{\rm H\alpha}$}
\newcommand{\WLi}{$EW_{\rm Li}$}
\newcommand{\gaia}{{\it Gaia}}
\newcommand{\tess}{TESS}
\newcommand{\prot}{$P_{\rm rot}$}
\newcommand{\rha}{$R^{'}_{\rm H\alpha}$}
\newcommand{\fha}{$F_{\rm H\alpha}$}
\newcommand{\erg}{erg\,cm$^{-2}$s$^{-1}$}

\newcommand{\eagles}{{\tt EAGLES}}

\definecolor{blu}{rgb}{0,0,1}
\definecolor{mag}{rgb}{1,0,1}
\definecolor{dgreen}{rgb}{0.1, 0.53, 0.22}
\definecolor{orange}{RGB}{255,127,0}

\begin{document}

\title{NGC\,1647: A young open cluster with a broad main sequence observed with LAMOST}

\author{A. Frasca\inst{1}\and 
	M.~Qin\inst{2,3,1}\and
        J. Alonso-Santiago\inst{1}\and
        G. Catanzaro\inst{1}\and
        J.~N. Fu\inst{2,3}\and
        J.~Y. Zhang\inst{4}\and
        A. Bragaglia\inst{5}
 }

\institute{INAF - Osservatorio Astrofisico di Catania, via S. Sofia, 78, 95123 Catania, Italy\\ \email{antonio.frasca@inaf.it}
\and
Institute for Frontiers in Astronomy and Astrophysics, Beijing Normal University, Beijing~102206, P.~R.~China\\ \email{mfqin@mail.bnu.edu.cn}
\and
School of Physics and Astronomy, Beijing Normal University, Beijing~100875, P.~R.~China
\and
Institut f\"ur Astronomie und Astrophysik, Eberhard Karls Universit\"at T\"ubingen, Sand 1, 72076 T\"ubingen, Germany
\and
INAF - Osservatorio di Astrofisica e Fisica dello Spazio, via P. Gobetti 93/3, 40129 Bologna, Italy
}

\date{Received 2 February 2026 / Accepted 7 April 2026}

\abstract 
{} 
{In this work we present the results of our analysis of medium-resolution LAMOST spectra of candidate members of the cluster NGC~1647 with the aim of determining the stellar parameters, 
activity level, lithium abundance, and to study the cluster properties.} 
{We used the code \rotfit\ to determine the atmospheric parameters (\teff, \logg, and \feh), radial velocity (\vrad), and projected rotation velocity (\vsini). Moreover, for 
solar-type and cooler stars (\teff$\le 6500$\,K), we calculated the \halpha\ and \ion{Li}{i}$\lambda$6708 net equivalent width by means of the subtraction of 
inactive photospheric templates. We determined the rotation periods for 160 stars by analyzing the available \tess\ photometry.}
{We derived \vrad, \vsini, and atmospheric parameters for 341 spectra of 155 stars, plus three additional bright targets with archival UVES spectra. Moreover, we found four 
double-lined spectroscopic systems for which we provide the radial velocities of the two components. The \vrad\ distribution of the cluster 
members peaks at $-$5.3\,\kms\ with a dispersion of 1.6\,\kms, while the average metallicity is \feh\,=\,$-$0.08$\pm$0.08\,dex, in line with previous determinations, which were based 
on only a handful stars. From the fitting of the spectral energy distribution of 160 likely members we 
infer the existence of a differential reddening across the cluster field with an average value of $A_V$=1.1\,mag. The $A_V$ values show a distinct correlation with the color offset 
from the lower boundary of the main sequence, as observed in the \gaia\ color-magnitude diagram; conversely, this offset appears to be uncorrelated with \vsini. 
These two findings confirm that differential reddening is the primary driver behind the observed extended Main-Sequence Turn-Off (eMSTO) in this cluster. 
The age of NGC 1647, obtained from the lithium abundance, is 203$\pm$27\,Myr, which is compatible with the values inferred from a gyrochronological approach and the 
isochrone fitting.}
{}


   \titlerunning{NGC\,1647 cluster as seen by LAMOST}
      \authorrunning{A. Frasca et al.}

\maketitle
\nolinenumbers

\section{Introduction}
\label{Sec:Intro}

The study of open star clusters (OCs) is fundamental for understanding the formation and evolution of the Galactic disk. Crucially, as groups of stars born simultaneously from the same 
progenitor cloud, OCs represent homogeneous samples in terms of both age and initial chemical composition (metallicity). This inherent uniformity makes them powerful tools for testing 
stellar evolutionary models, calibrating astrophysical distance scales, and mapping the chemical and structural properties of the Milky Way.

The advent of the $Gaia$ mission has revolutionized cluster studies by providing highly accurate astrometric and photometric data for billions of stars. The extreme precision of these 
data has enabled an unprecedentedly complete census of cluster membership and allowed for the definitive determination of cluster parameters, including distance, proper motion, and 
stellar density profiles. Recent efforts, notably by \citet{Cantat2018, Cantat2020} and \citet{Hunt2024}, have leveraged the massive $Gaia$ datasets to redefine the membership and 
fundamental properties of thousands of OCs across the Galaxy.

The high photometric precision provided by $Gaia$ also yields color-magnitude diagrams (CMDs) of exceptional detail. These CMDs reveal features such as the binary star sequence 
and the extended main-sequence turn-off (eMSTO) region of the main sequence (MS), the origin of which remains a subject of active debate in stellar astrophysics.

In this context, NGC\,1647 serves as an excellent case study. Located in the Taurus-Auriga region, its CMD exhibits a broad MS \citep{Zdanavicius2005}. While the eMSTO phenomenon 
can contribute to this feature, the cluster's position suggests that the observed broadness may be significantly influenced by differential extinction caused by intervening dust 
clouds in the Taurus region.
NGC 1647 is a poorly studied cluster with only a few astrometric and photometric studies dating back several decades \citep[e.g.,][]{1915ApJ....42..120S,1996A&AS..118..277G,Zdanavicius2005}. 
More recently, NGC\,1647 has been included in the large surveys that have automatically processed $Gaia$ data \citep[e.g.,][]{Cantat2020,Hunt2024}. Both studies place the cluster 
at around 600 pc but differ on its age (117--363\,Myr) and extinction ($A_V$=0.64--1.35 mag).

To address these issues and to derive better cluster parameters, such as the age and metallicity, spectra of a significant number of members in different regions of the CMD are needed. 
In this work, we conducted a comprehensive study of the cluster properties based on the characterization of the largest sample of members to date.
By combining mid-resolution spectroscopy ($R\simeq$7500) from LAMOST and $Gaia$~DR3 astrometric and photometric data we delved deeper into the identification of the cluster 
members and their individual properties, such as atmospheric parameters, activity level, and lithium abundance. This allowed us to examine cluster properties and to investigate the 
origin of the observed eMSTO and broad MS.
 
This paper is structured as follows. Section\,\ref{Sec:Data} describes the selection of cluster members and the extraction of LAMOST mid-resolution data. Section\,\ref{Sec:Analysis} 
details the methodology used for the spectroscopic determination of stellar parameters and extinction. Section\,\ref{Sec:Results} presents the refined cluster properties, including 
the radial velocity distribution of the cluster members, a revised age, and metallicity. Section 5 discusses the implications of these findings, particularly in the context of the 
eMSTO and differential extinction and, finally, we summarize our conclusions in Sect.\,\ref{Sec:Concl}.

\section{Observations and sample selection}
\label{Sec:Data}

\begin{figure*}[ht]
\begin{center}
\includegraphics[width=6.15cm,viewport= 30 0 380 330]{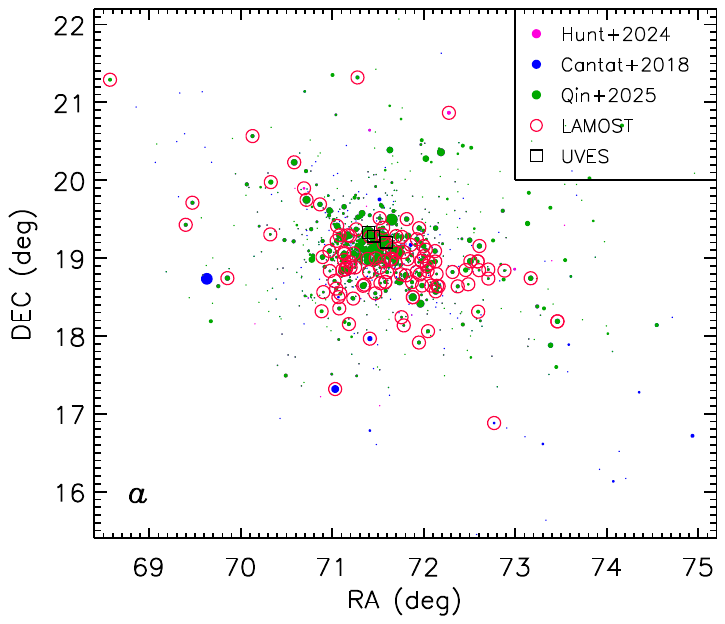}
\hspace{-.4cm}
\includegraphics[width=6.45cm,viewport= 10 0 380 330]{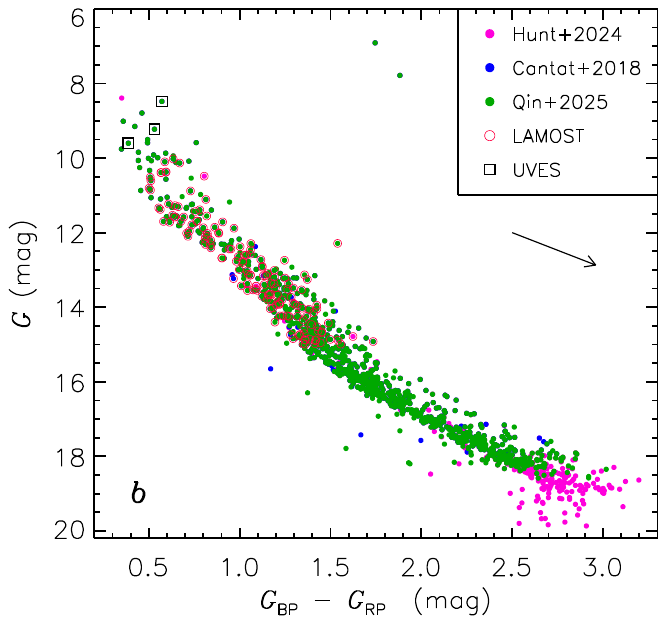}
\hspace{-.7cm}
\includegraphics[width=6.45cm,viewport= 10 0 380 330]{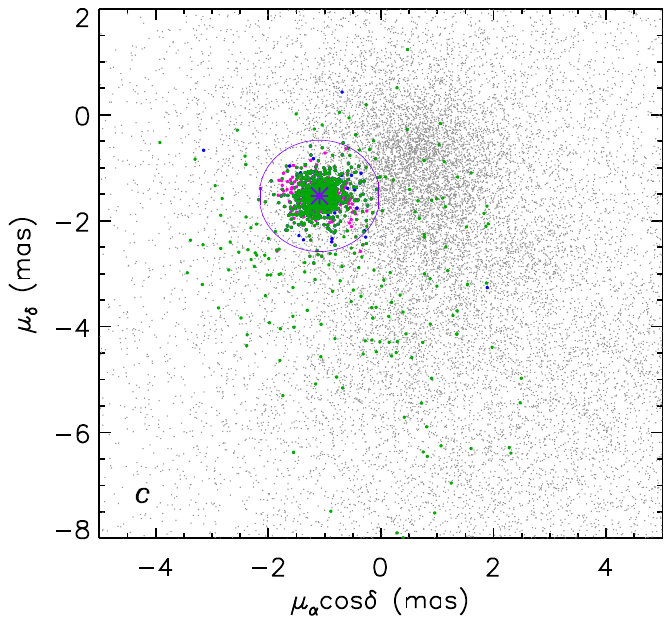}
\caption{Left panel: Spatial distribution of the NGC\,1647 members according to \citet[][magenta dots]{Hunt2024}, \citet[][blue dots]{Cantat2018}, and 
\citet[][green dots]{Qin2026}.
Symbol size scales with the $G$ magnitude.
The red circles highlight the data points that correspond to objects observed with LAMOST MRS. The black squares enclose the three stars with UVES archive observations. The meaning 
of the symbols is also indicated in the legend.
Middle panel: Color--magnitude diagram of the same sources. The arrow indicates the extinction vector for a value of $A_V=1.1$\,mag. Right panel: Proper motion diagram of all the 
\gaia\ DR3 sources with $G\leq 16$\,mag (small gray dots) in the field of NGC\,1647 (center coordinates RA(2000) = 71.5$^{\circ}$, DEC(2000) = +19$^{\circ}$, radius=3$^{\circ}$). 
The cluster members are highlighted with the same symbols as in the other panels (red circles have been omitted to avoid confusion).
The violet asterisk denotes the average proper motion of the cluster according to \citet{Hunt2024} and the violet ellipse is the 5$\sigma$ contour.
}
\label{Fig:Sky_CMD_PPM}
\end{center}
\end{figure*}

\subsection{Sample selection}
\label{Subsec:Sample}

To select the candidate members of this cluster, we used the catalog of \citet{Cantat2018}, who listed 645 candidate members selected on the basis of \gaia\ DR2 
astrometry and photometry (down to $G=18$\,mag) and  the application of an unsupervised membership assignment code ({\tt UPMASK}). 
Additionally, we used the new catalog of members of Galactic OCs compiled by \citet{Hunt2024}, who used a similar approach, but with a different algorithm, namely  the 
Hierarchical Density-Based Spatial Clustering of Applications with Noise ({\tt HDBSCAN}) that was applied to the \gaia\ DR3 data down to magnitude $G\approx 20$\,mag. 
They found 982 members, 623 of which are in common with \citet{Cantat2018}.
The higher number of sources found by \citet{Hunt2024}, apart from the different algorithm, is mostly the result of a larger sampled sky area and the deeper limiting magnitude. 

A list of 931 candidate members was compiled by \citet{Qin2026} who also applied the {\tt HDBSCAN} code to a sample of 40,070 \gaia\ sources selected 
around the center of NGC\,1647 defined by \citet{Cantat2018} as explained in their work. 
In the latter sample, 764 objects overlap with the catalog presented by \citet{Hunt2024} and 614 objects overlap with that of \citet{Cantat2018}. Furthermore, 610 stars are common to 
all three samples, which we will henceforth refer to simply as the {\it C-G}, {\it H\&R}, and {\it Qin} samples. We considered this subset of 610 stars (henceforth the `golden sample') 
to be bona fide candidates, electing not to use the membership probabilities, which are defined differently across the respective papers.

\subsection{LAMOST Spectroscopy}
\label{Subsec:Spectra}

Following a five-year low-resolution spectroscopic survey (LRS; $R_{\rm LRS} \simeq 1800$, $\lambda\in$\,[3000,9000]\,\AA, \citealt{zhao2012,luo2015}),
\lamost\ launched its medium-resolution spectroscopic survey (MRS; $R_{\rm MRS} \simeq 7500$) in September 2017 \citep{Liu2020}. The MRS covers the $\lambda\in$\,[4950,5350]\,\AA\ (blue) and 
$\lambda\in$\,[6300,6800]\,\AA\ (red) bands.

In this work, we utilized \lamost\ MRS DR12\footnote{\url{https://www.lamost.org/dr12/v1.0/}}, which comprises about 46 million spectra collected through June 2023. This data release is 
presently restricted to the Chinese astronomical community. 
The cross-match of our sample of members with the \lamost\ DR12 catalog, adopting a radius of 3$\farcs$7 on the basis of the fiber pointing precision (0$\farcs$4) and the 3$\farcs$3 
diameter of the fiber \citep[e.g.,][]{Zong2018}, produced 173 targets with a total of 428 MRS spectra, 
159 of which have spectra with a sufficient signal-to-noise ratio (S/N typically $\geq5$) in at least one arm that we considered worthy to be analyzed (red circles in 
Fig.~\ref{Fig:Sky_CMD_PPM}). 
In total we analyzed 345 \lamost\ MRS co-added spectra corresponding to 159 candidate members of NGC\,1647,  
114 of which are in the `golden sample' of bona fide members,
while 45 of them belong to two (or only one) of the {\it C-G}, {\it H\&R}, and {\it Qin} samples and are thus considered lower probability members. 
To estimate the potential contamination of the golden sample, we cross-checked the membership probabilities ($P$) from the three reference catalogs. Out of the 114 golden sample 
sources, 113 (99\%) have a membership probability $P>0.5$ in at least one catalog, while 80 sources (70\%) meet the more stringent criterion of $P>0.5$ in all three catalogs 
simultaneously. This indicates that the fraction of potential non-members is negligible (less than 1\%) under a relaxed threshold and remains well-constrained at $\approx 30$\,\% even 
when adopting the most restrictive membership criteria.
We chose to use co-added spectra, where each spectrum is the sum of all exposures obtained during a single observing night, to achieve the best possible S/N.

A precise evaluation of the spectral resolution, $R_{\lambda}$, during the observations of our targets is a necessary prerequisite to the subsequent analysis, especially for correctly 
measuring rotational broadening. To achieve this, we used the spectra of Th-Ar calibration lamps acquired during the science observations. For both arms we found values larger than the 
nominal $R_{\lambda}=7.500$: specifically, $R_{\rm blue}=8200\pm200$ and $R_{\rm red}=8400\pm300$. More details can be found
in Appendix\,\ref{Appendix:Resolution}.

\subsection{Archive spectra}
\label{Subsec:Obs_archive_spe}
Since the \lamost\ spectra do not cover the brightest MS targets, which are critical for constraining the eMSTO and the cluster age, we supplemented our dataset with archival data. 
We retrieved high-resolution spectra for three such stars: \object{HD~30123}, \object{HD~284839}, and \object{HD~284841}. 
These spectra were originally collected in February 2019 by \citet{Siebenmorgen2020}, under ESO program 0102.C-0040, with UVES at the VLT-U2 telescope, using the CD\#2 and CD\#4 
cross-dispersers and resolutions of about 65,000 and 74,500 for the blue and red arm, respectively. The aim of these observations was to investigate interstellar clouds. 
While \citet{Fitzpatrick2007} classified these targets as B8\,III or B9\,III, \citet{Siebenmorgen2020} assigned spectral types in the range B5--B8 and luminosity class V or IV. 
Notably, HD~30123, the second brightest MS star in the cluster, is identified as a Be star, exhibiting distinct core emission in both Balmer and Paschen lines. We used these spectra 
for determining the atmospheric parameters and radial and rotational velocities as for the stars with \lamost\ data.

\subsection{Photometry}
\label{Subsec:Obs_photo}

Time-series photometric observations form the foundation for studying stellar rotation periods. In NGC\,1647, a total of 112 members were observed by the {\it K2} mission 
\citep{Howell2014,Furlan2018}. 
However, because all stars with {\it K2} light curves also have observations from the Transiting Exoplanet Survey Satellite (\tess; \citealt{Ricker2015}), and since our primary goal 
was to determine reliable rotation periods for the largest number of targets as possible, we based our analysis exclusively on the {\tess\ photometry. In total, 623 candidate 
members possess \tess\ light curves. All of them were observed in Sectors~43 (September 16 to October 11, 2021) and~44 (October 12 to November 5, 2021), providing nearly continuous, 
high-quality photometry over approximately 55 days with a cadence of 600\,s. A subset of stars also received additional monitoring in Sectors~70 (September 20 to October 16, 2023) and~71 
(October 16 to November 11, 2023), where the exposure time was 200\,s. These additional data extend the observational baseline and enable independent confirmation of rotational 
signals when present. We retrieved the \tess\ Science Processing Operations Center (SPOC) light curves \citep{Caldwell2020} available in the Mikulski Archive for Space Telescopes
(MAST)\footnote{\url{https://mast.stsci.edu/portal/Mashup/Clients/Mast/Portal.html}} through the \texttt{Lightkurve} package \citep{Lightkurve2018}. The latter was also used for a 
quick look and data preparation for the following analysis, which was based on the Pre-search Data Conditioning Simple Aperture Photometry (PSDCSAP).

\section{Data analysis}
\label{Sec:Analysis}

The code \rotfit\ \citep[e.g.,][]{Frasca2006,Frasca2015} was used to measure \vrad, \vsini, and the atmospheric parameters (\teff, \logg, and \feh). The version of 
\rotfit\ developed for \lamost\ MRS was described by \citet{Frasca2022}, who analyzed thousands of MRS spectra in the \kepler\ field, and by \citet{Frasca2025}, who applied it 
to the MRS spectra of the Pleiades cluster members.
For objects cooler than about 7000\,K (henceforth `cool stars'), as judged from the spectrum appearance and the position on the CMD, ELODIE archive spectra ($R\simeq$42,000; 
\citealt{Moultaka2004}) of slowly rotating stars (\vsini\,$\leq3$\,\kms) with a low activity level (based on the residual flux in the H$\alpha$ core) were used as templates. 
This grid is composed of spectra of 388 different stars of FGKM spectral types (\teff$\leq$\,7000\,K), which sufficiently cover the space of parameters, especially at metallicity 
values around \feh=0\,dex. 
For stars hotter than or close to 7000\,K (hereafter `hot stars'), we used instead a grid composed of synthetic BT-Settl spectra \citep{Allard2014} with a solar metallicity, \teff\ 
in the range 6000--12,000\,K, and \logg\ between 3.0 and 5.0 (in steps of 0.5 dex). As the grid of the spectra is computed with steps of \feh\,=\,0.5\,dex, we used the value of \feh=0\,dex, 
which is the closest to the previously-derived cluster metallicity.
These stars are located in the \gaia\ CMD at a $G\le$\,12.9\,mag and $G_{\rm BP}-G_{\rm RP}$ color bluer than 1.12\,mag.
For a few stars close to the threshold for `hot stars' we used both template grids to check the consistency of the results. An example of the fitting results by the \rotfit\ code 
with both template grids is shown in Fig.\,\ref{Fig:ROTFIT_output}. 
For the cool stars we opted for real-star templates, because they reproduce more faithfully the photospheric features and they are best suited for the measure of H$\alpha$ chromospheric 
core emission and the \ion{Li}{i} absorption by the application photospheric spectral subtraction.

For the stars with more \lamost\ MRS spectra we computed, for each arm, the atmospheric parameters for all the spectra (excluding those with strong flaws) and 
averaged them using a variance-defined weight ($w_i=1/\sigma_i^2$), where $\sigma_i$ is the error of the given parameter in the $i$-th measure.  
Therefore, we obtained two sets of parameters per each star, one derived from the analysis of the blue-arm and the other of the red-arm spectra. 
The comparison of the results from the two arms was used to evaluate the accuracy of our measures in different regions of the parameter space (Sects.\,\ref{Subsec:RV} and \ref{Subsec:APs}). 
The final parameters, which are reported in Table~\ref{Tab:APs}, are the weighted mean of the values derived from each arm, using the variance-defined weights, as before. 
The uncertainties for the parameters listed in Table~\ref{Tab:APs} represent the maximum value between the standard error of the weighted mean and the weighted standard 
deviation derived from the measurements obtained from the blue- and red-arm spectra.

\setlength{\tabcolsep}{0.1cm}

\begin{table*}[ht]
  \caption{Stellar parameters of the investigated sources.}
\begin{center}
\begin{tabular}{llccclclcc}
\hline\hline
\noalign{\smallskip}
 Star$^a$  &  \gaia-DR3     & ~~~\teff  & \logg  &  \feh~~  &  \vsini~~  & $<$\vrad$>$  &  Sp.       & Sample$^b$  &  Remarks \\ 
 &    & ~~~(K) & (dex)  & (dex) &   (\kms) &  (\kms) & type &  &\\  
\noalign{\smallskip}
\hline
\noalign{\smallskip}
  HD~30123          &  \scriptsize{3410108311089573376} & 12500$\pm$200 & 3.65$\pm$0.05 &  0.00          &  60.0$\pm$2.0   &  -6.02$\pm$2.00 &  hot  &  CHQ  & UVES$^c$  \\
 HD~284839          &  \scriptsize{3409918164297368704} & 12000$\pm$250 & 3.70$\pm$0.10 &  0.00          &  140.0$\pm$30.0 & -13.00$\pm$5.00 &  hot  &  CHQ  & UVES$^c$  \\
 HD~284841          &  \scriptsize{3410107142858469248} & 11750$\pm$250 & 3.70$\pm$0.10 &  0.00          &  260.0$\pm$30.0 &  -3.62$\pm$5.00 &  hot  &  CHQ  & UVES$^c$  \\  
\scriptsize{J043417.11+211726.4} &   \scriptsize{144755932773284224} &  5816$\pm$ 45 & 4.38$\pm$0.07 &  0.03$\pm$0.07 &  $\leq$ 8.0     & -12.83$\pm$1.69 & G2IV  &    Q  & RVvar$^d$ \\
\scriptsize{J043735.67+192538.6} &  \scriptsize{3410572717313072768} &  5799$\pm$ 42 & 4.31$\pm$0.05 & -0.13$\pm$0.05 &  42.4$\pm$1.3   &  -3.36$\pm$0.37 &  G3V  &  CHQ  & ... \\
\scriptsize{J043753.27+194239.1} &  \scriptsize{3410615563906786048} &  5804$\pm$ 30 & 4.36$\pm$0.04 &  0.01$\pm$0.05 &   8.6$\pm$1.1   &  -4.85$\pm$1.02 &  G2V  &  CHQ  & ... \\
\noalign{\smallskip}
\hline
\end{tabular}
\end{center}
{\bf Notes.} The full table is available at the CDS. $^{(a)}$ \lamost\ designation or HD number.
$^{(b)}$ Subsample to which the target belongs: C\,=\,C-G; H\,=\,H\&R; Q\,=\,Qin. $^{(c)}$ UVES archival spectra. $^{(d)}$ `RVvar' indicates stars with genuine RV variations. 
\label{Tab:APs}
\end{table*}

We analyzed the three archival UVES spectra using a spectral-synthesis approach, since the combination of high effective temperature and rapid rotation results in a severe paucity 
of measurable metal lines. The synthetic spectra were computed in three steps. First, LTE model atmospheres were calculated with the ATLAS9 code \citep{1993ASPC...44...87K}. 
Second, the corresponding synthetic spectra were generated with the radiative-transfer code {\tt SYNTHE} \citep{1981SAOSR.391.....K}. Third, the best-fitting solution was obtained 
with ad hoc {\tt IDL} routines by minimizing the $\chi^2$ of the residuals between observed and synthetic spectra. The stellar parameters \teff\ and \logg\ were primarily constrained 
from the Balmer-line profiles in the blue UVES arm. The projected rotational velocity, $v \sin i$, was determined by fitting the \ion{Mg}{ii}\,$\lambda$4481\,\AA\ line, and all 
synthetic spectra were convolved with the instrumental line-spread function (set by the UVES resolving power) and with the rotational broadening kernel before comparison with the 
observations. Owing to the small number of iron lines and the strong rotational broadening, the metallicity could not be inferred. 
Therefore, we fixed it to \feh\,=\,0\,dex, consistently with the assumption adopted for the hot \lamost\ sources.

\subsection{Radial velocity}
\label{Subsec:RV}

\begin{figure}[htb]
\includegraphics[width=8.5cm,viewport= 0 0 520 530]{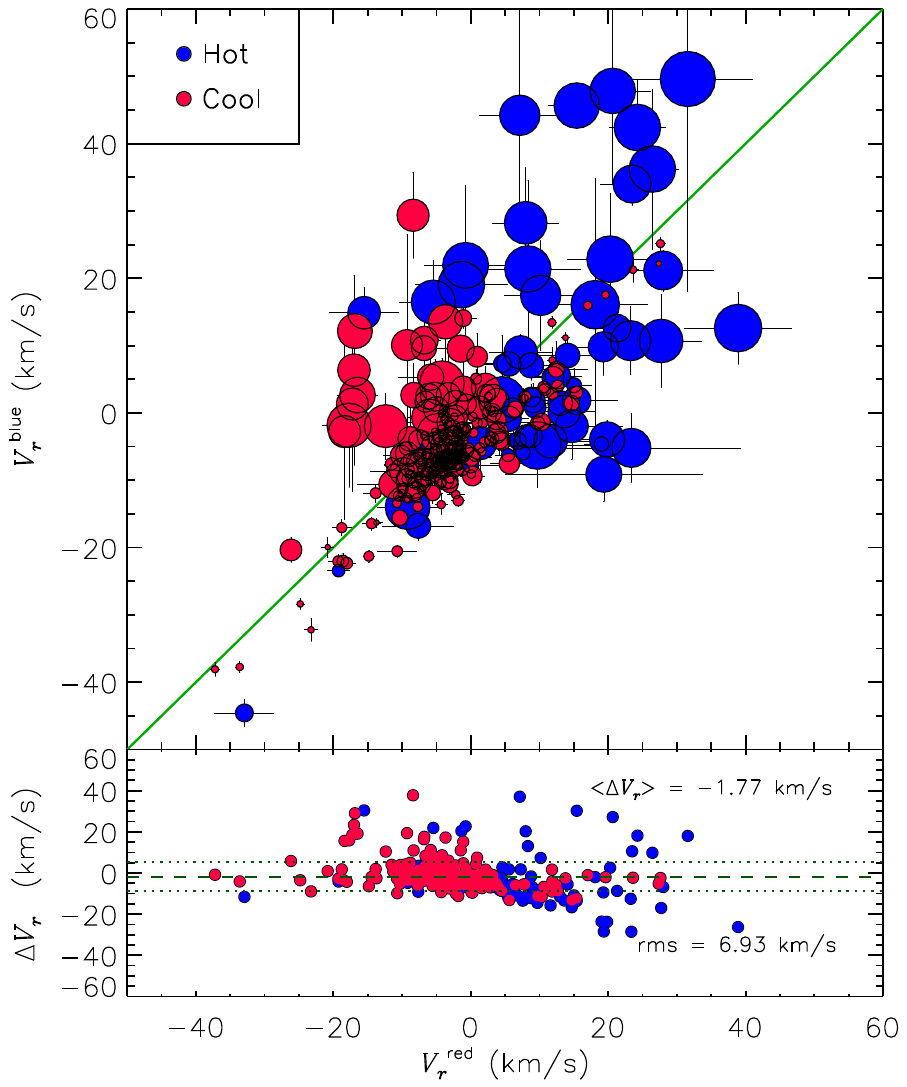}
\caption{Comparison between the radial velocities measured from blue- and red-arm \lamost\ MRS spectra (top panel). Blue and red symbols are used for 
the `hot' and `cool' stars, analyzed with BT-Settl and ELODIE grids of templates, respectively. 
The symbol size scales with the \vsini\ of the source.
The one-to-one relation is shown by the solid green line. 
The differences $\Delta$\vrad\,=\,$V_r^{\rm blue}-V_r^{\rm red}$ are displayed in the bottom panel along with their average, $<\Delta$\vrad$>$, and standard deviation, rms.
}
\label{Fig:RV_blue_red}
\end{figure}

The radial velocities measured on \lamost\ MRS spectra are affected by systematic offsets in different runs that are related to the wavelength calibration and are different 
for the blue and red arm of each spectrograph \citep[e.g.,][]{Zong2020ApJS..251...15Z,Frasca2022}.  
To correct for the radial velocity zero-point offsets, we cross-matched our sample with the official \lamost\ DR\,12 catalog and adopted the corresponding zero-point corrections 
for the red and blue arms provided therein. For spectra without available correction parameters (i.e., those with default values of $-9999$ in the \lamost\ DR12 catalog), we 
applied the typical offsets reported by \citet{Wang2019}, namely $+0.25$~\kms\ for the blue arm and $-0.09$~\kms\ for the red arm. 

The \vrad\ uncertainties derived from \rotfit\ and reported in Table~\ref{Tab:APs} range from 0.4 to 27.3\,\kms, with a median value of 2.2\,\kms. To investigate the precision of 
the \vrad\ measures and its dependence on other parameters, such as the \vsini, is to compare the results of the blue- and red-arm spectra, 
which are obtained simultaneously. This comparison is shown in Fig.~\ref{Fig:RV_blue_red}, where we used a symbol size proportional to \vsini\ and different colors for the `hot' 
and `cool' stars that were analyzed with different grids of templates, as explained in Sect.\,\ref{Sec:Analysis}. The largest scatter around the one-to-one relation is displayed 
by the fastest rotators, which are mostly hot stars. 
The root-mean-square (rms) of the difference $\Delta$\vrad\,=\,\vrad$^{\rm blue}$ -- \vrad$^{\rm red}$ is about 6.9\,\kms, which reduces to about 3.6\,\kms\ if we exclude the 
stars with \vsini$\geq 70$\,\kms. 
By dividing this value by $\sqrt2$, as it includes the uncertainties of the \vrad\ in the two arms, we get an estimate of 2.5\,\kms\ for the average precision of the \vrad\ 
measurements, which is close to that found by us for the \lamost\ MRS observations of the Pleiades \citep{Frasca2025}.

We performed a search for sources displaying intrinsic radial velocity variability -- indicative of binarity or pulsations -- by analyzing objects for which multiple spectra 
were acquired on separate nights. To identify statistically significant variations, we computed the reduced $\chi^2$ statistic and the corresponding probability $P(\chi^2)$ that the 
\vrad\ scatter is merely due to random chance \citep[e.g.,][]{Press1992}. We assigned the `RVvar' flag in Table~\ref{Tab:APs} to the 31 sources where $P(\chi^2)<0.05$, indicating 
significant \vrad\ variation. For these candidates, we have between two and five spectra per source, with a median of three.
We found no objects previously classified as single-lined spectroscopic binaries (SB1s) in our sample, with the exception of \gaia\,DR3\,3409916343231260032, which is classified as 
an SB1 in the {\it Gaia DR3. Non-single stars} catalog \citep{GAIA-SB1}. However, this source was not analyzed by us for the atmospheric parameters, due to the low S/N of the 
single \lamost\ MRS spectrum we possess. The \vrad\ derived from the blue arm, $-25.6$\,$\pm$\,1.7\,\kms, is significantly different from the cluster average, as expected 
for an SB1 system.
Furthermore, for four objects (five spectra in total), we observed two CCF peaks above the noise that were deemed significant. These objects were classified as new double-lined 
spectroscopic binaries (SB2s). Since the parameters derived for these objects are unreliable, they are not included in Table~\ref{Tab:APs}; instead, we report the measured \vrad\ 
values for both components of these new SB2s in Table~\ref{Tab:SB2}, where \vrad$_1$ corresponds to the component with the highest CCF peak.

For a quality check, we compared in Fig.~\ref{Fig:Comp_RV_Gaia} the average \vrad\ values of the single-lined sources (Table~\ref{Tab:APs}) with those contained in the Gaia DR3 
catalog, which were derived using the RVS spectrograph. Many targets follow a one-to-one relation and cluster around the mean radial velocity of NGC1647, with some sources, including 
a few among those classified by us as 'RVvar', exhibiting significant scatter.

\subsection{Atmospheric parameters and projected rotation velocity}
\label{Subsec:APs}

\begin{figure*}[htb]
\includegraphics[width=18cm]{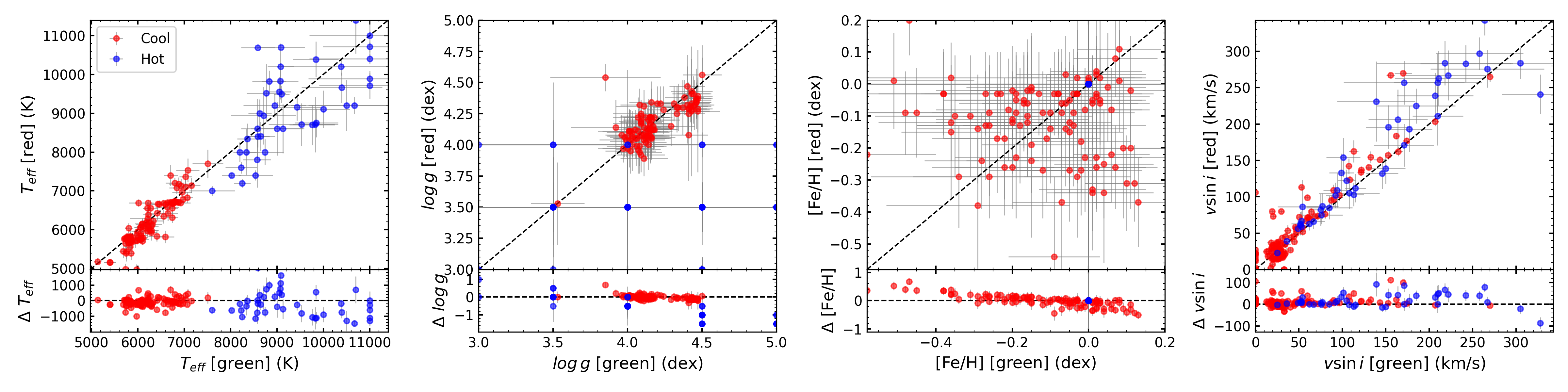}
\caption{Comparison of the atmospheric parameters derived from the blue- and red-arm LAMOST MRS spectra using ROTFIT. Panels, from left to right, show the effective temperature 
(\teff), surface gravity (\logg), metallicity (\feh), and projected rotation velocity (\vsini). The symbols are consistent with those presented in Fig.~\ref{Fig:RV_blue_red}. 
Note the significantly larger rms dispersion observed for the \teff\ of the hot stars (blue data points).}
\label{Fig:Comp_blue_red}
\end{figure*}

Comparing the blue- and red-arm values for the other stellar parameters (\teff, \logg, \feh, and \vsini) allows us to obtain reliable estimates of their errors in different regions 
of the parameter space and to highlight any specific issues encountered during their determination (Fig.\,\ref{Fig:Comp_blue_red}).
As explained previously, we generated two sets of parameters for each source -- one from the analysis of the blue-arm spectrum and one from the red-arm spectrum -- which we can 
compare, as we did for the radial velocities.

Regarding the effective temperature, the measured values generally follow the one-to-one relation, although a significantly larger scatter is apparent for the `hot' stars compared 
to the `cool' ones. This is likely due to the characteristics of early-type spectra, which feature fewer and shallower absorption lines, particularly in the red spectral range, combined 
with higher average \vsini\ values that further blend the available diagnostic features.
Specifically, the average difference between blue-arm and red-arm effective temperature is $\Delta T_{\rm eff}=T_{\rm eff}^{\rm blue}-T_{\rm eff}^{\rm red}=-88$\,K for the cool stars 
and $-162$\,K for the hot stars. 
The rms deviation is $\approx$\,250\,K for the cool stars, while it increases to about 850\,K for the hot ones. This larger dispersion for the hot targets is consistent with the 
average errors evaluated by \rotfit, which scale with the \teff, yielding median values of 55\,K for the `cool' stars and 210\,K for the `hot' ones.

The remaining parameters (\logg, \feh, and \vsini) exhibit very good agreement for the cool spectra (ELODIE templates) between the two independent arm analyses, with very small average 
offsets and rms dispersions of approximately 0.15 dex, 0.18 dex, and 23 \kms, respectively.
For the hot stars, the gravity values are clustered around the available model steps (\logg=3.5\,dex, 4.0\,dex, and 4.5\,dex). Furthermore, the single data point observed in the \feh\ plot 
is a direct result of our choice to use only solar metallicity BT-Settl spectra.

The scatter on \vsini\ is significantly reduced for sources with \vsini\,$<$\,100\,\kms, dropping to about 15\,\kms. 
Notably, the \vsini\ measurements appear equally robust for both hot and cool spectra, as illustrated in the right panel of Fig.\,\ref{Fig:Comp_blue_red} and confirmed by a similar 
overall rms scatter ($\approx 30$\,\kms).
As previously reported by \citet{Frasca2022}, the resolution and sampling of the LAMOST MRS spectra prevent reliable measurements of \vsini\ values below 8\,\kms. Therefore, any 
measurement smaller than this limit in at least one arm must be considered a non-detection. Accordingly, we treated the final \vsini\ values (the weighted averages of the blue and red 
arms) that are less than 8\,\kms\ as upper limits and flagged them appropriately in Table~\ref{Tab:APs}.

A statistically significant comparison between the parameters measured by us using the \lamost\ MRS spectra and values available in the literature, which could serve as a quality 
control for our data, cannot be performed. In fact, there are only seven stars observed within the GALactic Archaeology with HERMES (GALAH) Survey \citep{Buder2021} and six stars 
observed by APOGEE \citep{Abdurro2022} that overlap with those observed by \lamost\ MRS; the majority of these are cool objects.
The temperature values for these few overlapping sources show a substantially good agreement {(rms =337\,K, excluding the three hot sources observed by APOGEE). As regards the 
metallicity, the values from the literature are scattered around \feh\,=\,0\,dex with a dispersion much larger than our values.

Accurate parameter values were derived by \citet{Carrera2022} for the only two giant stars in the cluster using high-resolution spectra, but we do not possess \lamost\ MRS spectra for 
these sources.
Nonetheless, the parameters derived from those high-resolution data by \citet{Carrera2022} have been utilized in the subsequent analysis.

\section{Results}\label{Sec:Results}

\subsection{Radial velocity distribution}\label{Subsec:RV_distr}

\begin{figure}[htb]
\includegraphics[width=9cm,viewport= 30 10 450 280]{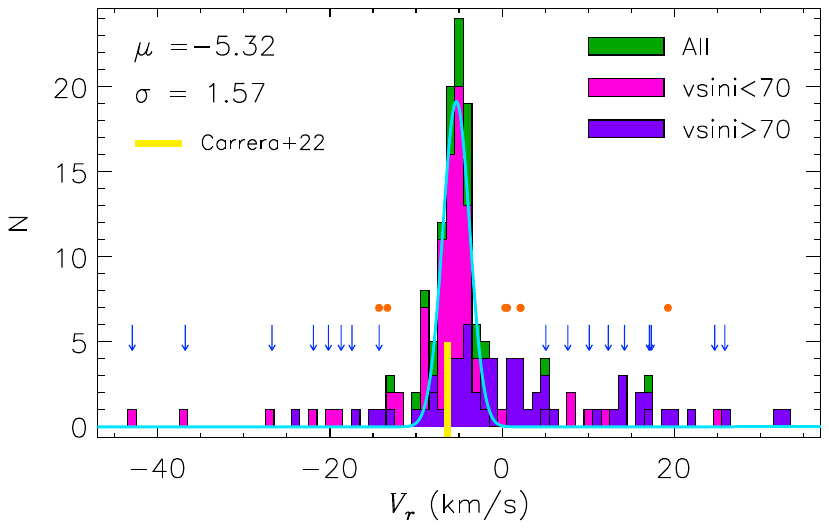}
\caption{Radial velocity distribution obtained from all analyzed \lamost\ MRS spectra of the cluster members (green histogram). Separate distributions are shown for the slowly rotating 
and fast rotating stars, as indicated in the legend. The best-fit Gaussian function applied specifically to the ``slowly-rotating stars'' histogram is overplotted as a solid cyan line; its 
center ($\mu$) and dispersion ($\sigma$) are also marked. The cluster \vrad\ measured by \citet{Carrera2022} is indicated by the vertical yellow bar. Downward blue arrows highlight 
lower-probability targets (those not included in the `golden sample') that lie outside the 5$\sigma$ limit of the Gaussian center. Within this same velocity range, orange dots denote the mean 
\vrad\ of sources exhibiting variable radial velocity.
}
\label{Fig:RV_distr}
\end{figure}

We have \vrad\ values for 158 single-lined sources (including the three MS stars with UVES spectra). For each of those with multiple \lamost\ observations we calculated the weighted 
average of the individual values (with $w=1/\sigma_{V_r}^2$ as the weight), which are also reported in Table\,\ref{Tab:APs} along with the error, which is the largest between the 
standard error and the weighted standard deviation of individual values. With these average radial velocities, we built the \vrad\ distribution, which is shown as a green histogram 
in Fig.~\ref{Fig:RV_distr}. 
The center $\mu=-5.18$\,\kms\ and the dispersion $\sigma=1.65$\,\kms\ of this distribution were found by means of a Gaussian fitting.  
We note that the intrinsic dispersion of the cluster RV distribution is narrower than the median formal uncertainty derived from \rotfit\ ($\approx 2.2$\,\kms). 
This suggests that our formal error estimates are likely conservative, particularly for stars with high signal-to-noise spectra and low rotation.

Since a relevant fraction of the selected candidates is composed of warm (\teff$\geq 7000$\,K) and fast rotating stars, we distinguished the  ``slowly'' (\vsini$\leq 70$\,\kms) and 
the ``fast'' rotating (\vsini$> 70$\,\kms) stars, for which the measure of \vrad\ is less accurate. We have overplotted their \vrad\ distributions with different colors in 
Fig.\,\ref{Fig:RV_distr}. 
As apparent, the fast rotating stars display a flatter and more scattered distribution compared to the slowly rotating ones. The distribution of the  latter subsample (94 sources) is 
similar to that of the full sample and is fitted  by a Gaussian with $\mu=-5.32$\,\kms\ and $\sigma=1.57$\,\kms. We consider the latter values as more representative for the average \vrad\  
of the cluster and its dispersion, although the inclusion of the fast-rotating stars does not significantly affect the result. 

Some of the values significantly far from the Gaussian center (by more than 5$\sigma$) may be partially attributed to single-lined spectroscopic binaries (SB1s) or pulsating stars. 
In Fig.~\ref{Fig:RV_distr}, we highlight potential SB1 systems (those labeled as `RVvar' in Table~\ref{Tab:APs}) whose average radial velocity falls within this range using orange 
dots.
The extended tail of the \vrad\ distribution may also be attributed to non-member contaminants or a stellar population characterized by a more dispersed kinematic distribution. Notably, this 
tail contains numerous stars excluded from the `golden sample' of high-probability members; these objects are highlighted with blue downward arrows in Fig.\,\ref{Fig:RV_distr}.

\subsection{Cluster metallicity}\label{Subsec:metal}

Similar to the other parameters, the final metallicity listed in Table~\ref{Tab:APs} is the weighted average of the values derived from the blue and red arms. The \feh\ distribution 
for the `cool stars' is shown in Fig.\,\ref{Fig:histo_feh_comb}, distinguishing the stars in the `golden sample' with a different color histogram. 
However, the two distributions are basically the same and are both centered at slightly under-solar values. By fitting them with a Gaussian, we derive an 
average cluster metallicity of \feh\,=\,$-0.12\pm0.11$\,dex, where the Gaussian dispersion ($\sigma$) is taken as the cluster metallicity uncertainty.
If we calculate instead the weighted average and standard deviation of the metallicity values we find $<$\feh$> = -0.08\pm0.08$\,dex for both the full and `golden' sample. 
These values are in good agreement with the robust metallicity determination made by \citet{Carrera2022} for the two giant member stars, which was based on high-resolution spectra.

\begin{figure}[htb]
\includegraphics[width=9cm,viewport= 30 10 450 280]{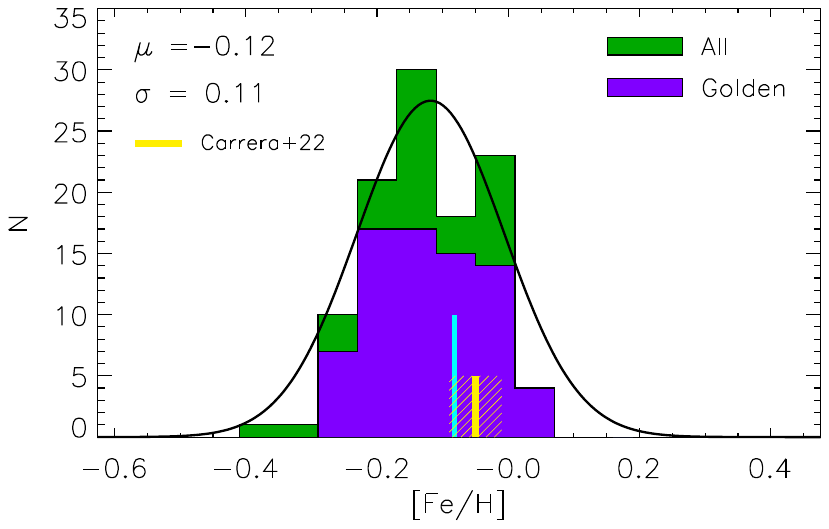}
\caption{Metallicity (\feh) distribution for all the `cool stars' of NGC\,1647 (green histogram) and for the high-probability `golden sample' (purple histogram). 
The best-fit Gaussian function is overlaid as a solid black line; its center ($\mu$) and dispersion ($\sigma$) are also marked. The cyan line denotes the weighted average of our \feh\ measures. 
The weighted-average metallicity measured by \citet{Carrera2022} for the two giants of the cluster is indicated by the vertical yellow bar, with the associated uncertainty represented by the 
hatched histogram. }
\label{Fig:histo_feh_comb}
\end{figure}

\subsection{SED Fitting: reddening and luminosities}\label{Subsec:redd}

To determine the interstellar extinction ($A_V$) and luminosity ($L$) for our sources, we employed the spectral energy distribution (SED) fitting method. We constructed the corresponding 
SEDs using publicly available optical and NIR photometric data and fitted them with BT-Settl synthetic spectra \citep{Allard2014}. 
Specifically, for the majority of sources, we utilized optical $BVg'r'i'$ photometry from the APASS catalog \citep{APASS} and NIR data from 2MASS \citep{2MASS}. Additionally, 
mid-infrared data from the WISE catalog \citep{WISE} were retrieved to identify potential mid-IR excesses and for visualization purposes. These data were excluded from the actual SED fit, 
which was restricted to the $BVg'r'i'JHKs$ bands. No sources exhibiting mid-IR excess were found. For the few sources lacking APASS entries, we substituted the optical data with 
Sloan photometry from the Pan-STARRS catalog \citep{Pan-STARRS}.

For each target, we fixed the distance using its $Gaia$-DR3 parallax and adopted the atmospheric parameters (\teff\ and \logg) derived in Sect.~\ref{Sec:Analysis}. We included the two 
giants, whose parameters were measured by \citet{Carrera2022}. This fitting method left the stellar radius ($R$) and $A_V$ as free parameters. These variables were determined via 
$\chi^2$ minimization, and the stellar luminosity was subsequently calculated as $L$=4\,$\pi$\,$R^2$\,$\sigma$\,$T_{\textrm{eff}}^4$. Two examples of the fitting procedure for a hot and 
a cool stars are presented in Figs.~\ref{Fig:sed1} and \ref{Fig:sed2}, respectively.
Uncertainties for $A_V$ and $R$ were derived from the 1$\sigma$ confidence level of the $\chi^2$ maps. To account for the \teff--$A_V$ degeneracy, we incorporated the uncertainty in 
\teff\ by recalculating the best-fit values at its upper and lower 1$\sigma$ limits.

\begin{figure}   
\centering            
\includegraphics[width=\columnwidth,viewport= 30 10 450 280]{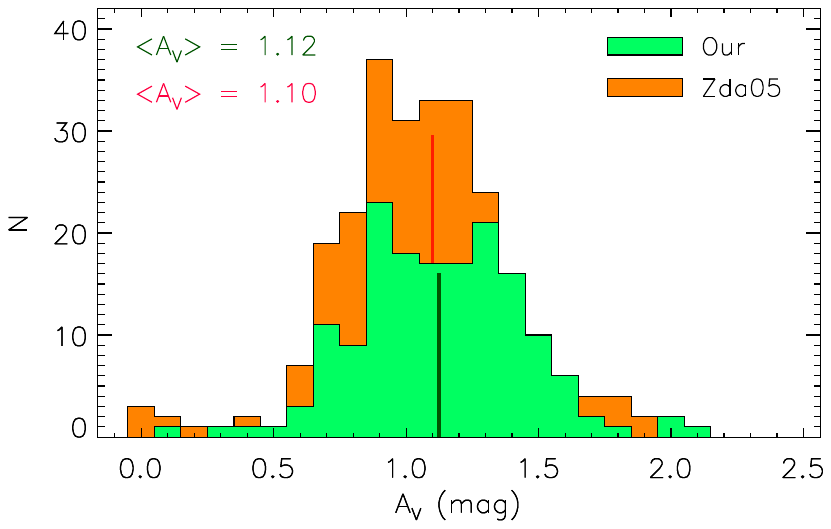}      
\caption{Extinction distribution for all the NGC\,1647 members with \lamost\ data (green histogram) and for the stars analyzed by \citet{Zdanavicius2005} (Zda05, orange histogram). 
The average $A_V$ values from both datasets are marked with vertical solid lines and reported with the corresponding color in the upper left corner of the box.}   
\label{Fig:A_V_distr} 
\end{figure}

The $A_V$ values determined by us via the SED fitting of the 160 sources with known parameters (including the three MS stars with UVES spectra and the two giants) range from 
about 0.3 to 2.1 mag and  are reported in Table~\ref{Tab:SED}. 
The $A_V$ distribution is displayed in Fig.~\ref{Fig:A_V_distr} along with that of the $A_V$ values measured by \citet{Zdanavicius2005}, which span the same range. The weighted average 
extinction from our measures is $A_V$=1.12\,$\pm$\,0.30\,mag, where the error represents the weighted standard deviation. The same average value, $A_V$=1.10\,$\pm$\,0.33\,mag, is found 
from the photometric determination by \citet{Zdanavicius2005}.
Assuming a standard reddening law with $R_V$=3.1, this corresponds to $E(B-V)=0.35\pm0.11$\,mag. For 43 sources we have $A_V$ measured both in this work and from \citet{Zdanavicius2005}. 
As shown in Fig.\,\ref{Fig:Av_comp}, these $A_V$ values are well correlated with each other, with a Pearson's correlation coefficient $\rho=0.677$. This reinforces the validity of our 
analysis. 

As a further test of the reliability of our extinction measurements, we have overlaid the stellar positions in Fig.~\ref{Fig:A_V_IRAS} on an IRAS 100\,$\mu$m emission map, displayed 
in grayscale with inverted intensity levels. This map effectively traces the dust column density along the line of sight. The symbols are color-coded by extinction value, demonstrating 
a strong spatial correlation between the measured $A_V$ and the 100\,$\mu$m thermal emission.

\begin{figure}   
\centering            
\includegraphics[width=\columnwidth,viewport= 0 0 410 390]{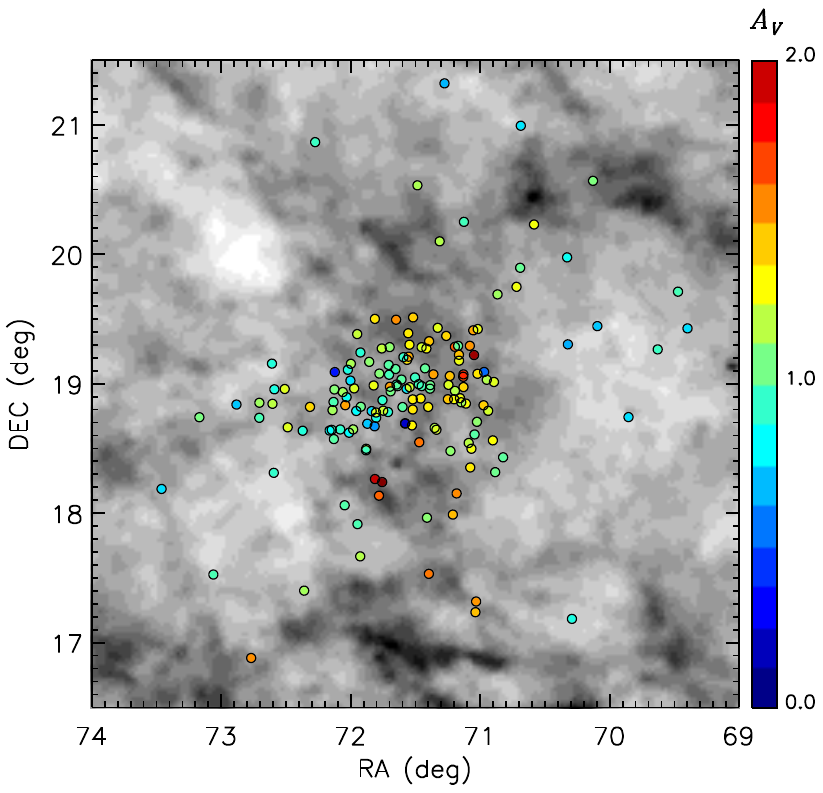}      
\caption{Spatial distribution of the stars in our sample. Symbols are color-coded according to their individual $A_V$ values. The grayscale image in the background is the 100\,$\mu$m 
IRAS flux map. }   
\label{Fig:A_V_IRAS} 
\end{figure}

\subsection{Hertzsprung-Russell diagram}\label{Subsec:HR}

An additional tool used to validate our results and infer the cluster age is the Hertzsprung-Russell diagram, presented in Fig.\,\ref{Fig:HR} alongside PARSEC isochrones 
\citep{2022A&A...665A.126N} for \feh\,=\,$-0.08$\,dex at four different ages. This diagram plots intrinsic (unreddened) quantities, as extinction was accounted for during the luminosity 
calculations.

As shown in the figure, most targets are located in the region where the MS isochrones overlap; consequently, the MS targets observed with \lamost\ MRS alone provide limited constraints 
on the cluster age. In this regard, the parameters of the two giants and the bright MS stars observed with UVES are particularly useful. Their positions in the Hertzsprung-Russell 
diagram allow us to exclude 
ages significantly younger than 150\,Myr or older than 200\,Myr. This age range is consistent with both previous determinations and the age inferred by us from the lithium depletion 
pattern (see Sect.\,\ref{Subsec:chrom_lithium}). Furthermore, this result is in good agreement with the asteroseismic age derived by \citet{Qin2026}.

In Fig.\,\ref{Fig:HR}, we have enclosed in green squares the stars marked with blue arrows in Fig.\,\ref{Fig:RV_distr}. These represent objects that do not belong to the `golden sample' 
and have \vrad\ deviating by more than 5$\sigma$ from the cluster mean ($-5.32$\,\kms). The most striking case is \gaia\,DR3\,3406926251422646400, which lies well above the MS and is 
identified as a subgiant (K0\,IV, \teff\,=\,5153\,K, and \logg\,=\,3.53\,dex) unrelated to the cluster.

\begin{figure}
  \centering
  \hspace{-.5cm}
  \includegraphics[width=\columnwidth,viewport= 20 0 500 360]{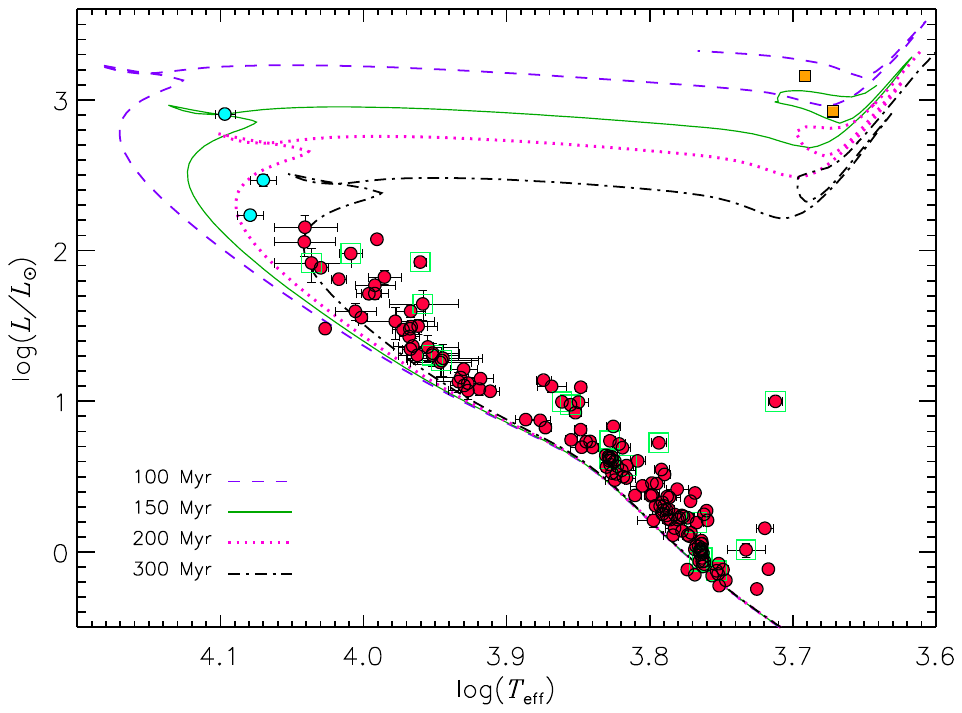}  
  \caption{HR diagram for NGC\,1647. The MS stars are denoted with dots (red for  \lamost\ MRS and cyan for UVES spectra). The orange squares represent the two giants studied by 
  \citet{Carrera2022}. PARSEC isochrones \citep{2022A&A...665A.126N} at 100, 150, 200, and 300\, Myr for models with the cluster metallicity (\feh\,=\,$-0.08$\,dex) are overlaid. 
  Green squares enclose the lower probability members indicated by arrows in Fig.\,\ref{Fig:RV_distr}.} 
  \label{Fig:HR} 
\end{figure}       
       
\subsection{Chromospheric emission and lithium abundance}
\label{Subsec:chrom_lithium}

\renewcommand{\tabcolsep}{0.1cm}

\begin{table*}[ht]
\caption{Activity indicators: net H$\alpha$ equivalent width (\Whalpha), line flux (\fha), and luminosity ratio (\rha\,=\,$L_{\rm H\alpha}/L_{\rm bol}$).}
\begin{center}
\begin{tabular}{cclccccc}
\hline\hline
\noalign{\smallskip}
Star$^a$ &  \gaia-DR3  & ~~~~\teff   & \logg  & \Whalpha  &  \fha   &  $\log$(\rha)    & Sample$^b$ \\  
   &  &  ~~~~(K) & (dex) & (\AA) & (\erg) & (dex) & \\  
\noalign{\smallskip}
\hline
\noalign{\smallskip}
 J043417.11+211726.4  &   144755932773284224  &  5816$\pm$45  &  4.38$\pm$0.07 &  0.08$\pm$0.01 &  6.302e+05$\pm$8.478e+04 &  -5.01$\pm$0.06  &     Q  \\ 
 J043735.67+192538.6  &  3410572717313072768  &  5799$\pm$42  &  4.31$\pm$0.05 &  0.11$\pm$0.02 &  8.224e+05$\pm$1.792e+05 &  -4.89$\pm$0.10  &   CHQ  \\ 
 J043753.27+194239.1  &  3410615563906786048  &  5804$\pm$30  &  4.36$\pm$0.04 &  0.25$\pm$0.02 &  1.867e+06$\pm$1.751e+05 &  -4.54$\pm$0.04  &   CHQ  \\ 
\noalign{\smallskip}
\hline
\end{tabular}
\end{center}
{\bf Notes.} The full table is available at the CDS. $^{(a)}$ \lamost\ designation. $^{(b)}$ Subsample to which the target belongs. 
\label{Tab:Halpha}
\end{table*}

For stars later than about F5 spectral type, both chromospheric emission (traced by the Balmer H$\alpha$ line in the \lamost\ MRS spectra) and lithium absorption are age-dependent 
parameters \citep[see, e.g.,][and references therein]{Jeffries2014, Frasca2018}.
Since chromospheric emission in \halpha\ can only show up as a small filling of the line core, which depends on the star's activity level and photospheric flux, removal of the 
underlying photospheric spectrum is mandatory. To achieve this, we subtracted a non-active, lithium-poor template spectrum that best matches the final atmospheric parameters 
from each \lamost\ red-arm spectrum. This template was aligned to the \vrad\ of the target, rotationally broadened by convolution with a rotational profile corresponding to the measured 
\vsini, and resampled onto the spectral points of the target spectrum.

The `emission' \halpha\ equivalent width (\Whalpha) was then measured by integrating the residual emission profile, as also shown in \citet{Frasca2022}. We adopted the convention 
of \Whalpha$>0$ for excess emission.
We excluded the SB2 systems from this analysis and retained only the \Whalpha\ values that were significantly larger than zero -- that is, those greater than or equal to their 
respective errors (70 sources). For the remaining 15 single-lined sources, the error was adopted as the upper limit of the measurement. These data are reported in Table~\ref{Tab:Halpha}, 
where we list the weighted mean of the values measured in the individual spectra for stars with multiple observations.
The equivalent width of a chromospheric line is generally not the optimal diagnostic for magnetic activity. More accurate indicators of chromospheric activity are the line flux in 
units of stellar surface (\fha) and the ratio between the line luminosity and bolometric luminosity (\rha). These values were evaluated according to Eqs.\,2 and 3 of \citet{Frasca2022} 
and are also reported in Table~\ref{Tab:Halpha}.

\begin{figure}[!t]
\hspace{0.5cm}
\includegraphics[width=9.2cm,viewport= 30 10 440 400]{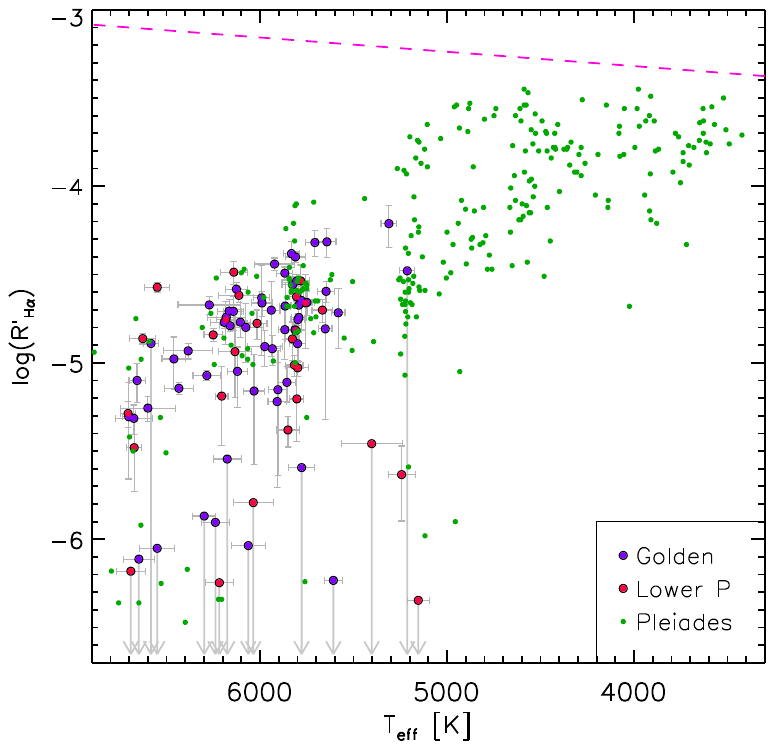}    
\caption{H$\alpha$ luminosity ratio (\rha) versus effective temperature (\teff). The plot distinguishes between the `golden sample' of likely members and the lower-probability 
targets using different colors, as specified in the legend. Upper limits on \rha\ are indicated by downward gray arrows. The straight dashed line represents the empirical boundary 
derived by \citet{Frasca2015} separating purely chromospheric sources from accreting objects. The \rha\ values for members of the Pleiades cluster from \citet{Frasca2025} are 
overplotted as small green dots for comparison.}  
\label{Fig:indicator}
\end{figure}

The H$\alpha$ luminosity ratio, \rha, is plotted as a function of \teff\ in Fig.~\ref{Fig:indicator}. This plot also displays the dividing line, empirically determined by 
\citet{Frasca2015}, which separates chromospherically active stars from objects still undergoing mass accretion. This line is situated close to the level of saturated chromospheric 
activity, defined as $\log$\rha\,$\simeq-3.3$ by \citet{Barrado2003}.
The \rha\ values measured in our study place all NGC\,1647 members firmly within the region of chromospherically active sources, largely occupying the same domain as the Pleiades 
members \citep{Frasca2025}. The only apparent difference between the two clusters in this diagram is the lack of MS stars cooler than about 5200\,K for NGC\,1647. This is the results 
of the larger distance and extinction of NGC 1647, which limits our ability to observe these fainter, cooler stars. This prevents us to study the saturated regime in this cluster, 
as we did in \citet{Frasca2025}, and to use this diagram as an age proxy. 

The equivalent width of the \ion{Li}{i} $\lambda$6708\,\AA\  absorption line (\WLi, defined as positive for absortpion) was also measured in the residual (subtracted) spectra, where blends with nearby photospheric 
lines had effectively been removed. 
This procedure ensures a better measure of \WLi\ and a reliable estimate of its error. The error was calculated as $\sigma_{EW_{\rm Li}} = D\cdot\sqrt{w\Delta\lambda}$, where $D$ is 
the average dispersion of the flux values in the residual spectrum on both sides of the line (with $D\simeq \frac{1}{S/N}$), $w$ is the integration width in wavelength units, and 
$\Delta\lambda$ (=0.15\,\AA) is the pixel size in wavelength units. This expression is consistent with the formula proposed by \citet{Cayrel1988}.

We successfully detected the lithium line (\WLi\,$> \sigma_{EW_{\rm Li}}$ in at least one spectrum) for 85 objects, while for 2 objects we could only determine an upper limit. 
We did not measure \WLi\ for the SB2 systems. For stars with time-series spectra, we calculated the weighted mean of the individual \WLi\ values and adopted the weighted standard 
deviation or the standard error of the weighted mean (whichever was greater) as the error estimate. For objects with only non-detections across all their spectra, we adopted the 
lowest upper limit. These final values are quoted in Table~\ref{Tab:lithium}.

\renewcommand{\tabcolsep}{0.1cm}

\begin{table*}[ht]
  \caption{Lithium equivalent widths (\WLi) and abundances ($A$(Li)).}
\begin{center}
\begin{tabular}{ccccccccc}
\hline\hline
\noalign{\smallskip}
 Star$^a$ &  \gaia-DR3  & RA     & DEC    & \teff & \logg & \WLi   &  $A$(Li)  & Sample$^b$ \\  
      &             &(J2000) &(J2000) & (K)   & (dex) & (m\AA) & (dex)     &                  \\  
\noalign{\smallskip}
\hline
\noalign{\smallskip}
  J043417.11+211726.4  &   144755932773284224 &  68.571305 &  21.290689  &  5816$\pm$45  &  4.38$\pm$0.07 &   30$\pm$5   &  2.13$\pm$0.11 &	Q \\
  J043735.67+192538.6  &  3410572717313072768 &  69.398650 &  19.427395  &  5799$\pm$42  &  4.31$\pm$0.05 &  105$\pm$22  &  2.73$\pm$0.14 &   CHQ \\
  J043753.27+194239.1  &  3410615563906786048 &  69.471989 &  19.710865  &  5804$\pm$30  &  4.36$\pm$0.04 &  131$\pm$18  &  2.85$\pm$0.10 &   CHQ \\
\noalign{\smallskip}
\hline
\end{tabular}
\end{center}
{\bf Notes.} The full table is available at the CDS. $^{(a)}$ \lamost\ designation. $^{(b)}$ Subsample to which the target belongs. 
\label{Tab:lithium}
\end{table*}

Lithium is a fragile element that is burned in stellar interiors at temperatures as low as 2.5$\times10^6$\,K. It is progressively depleted from the stellar atmosphere as internal 
mixing reaches the Li-burning layer, with the depletion rate dependent on the star's internal structure (i.e., stellar mass).
Consequently, Li abundance serves as a robust age proxy for stars cooler than about 6500\,K. 
We derived the lithium abundance, $A$(Li), from our values of \teff, \logg, and \WLi\ by interpolating the curves of growth presented by \citet{Lind2009}. These curves span 
the \teff\ range 4000--8000\,K and \logg\ from 1.0 to 5.0\,dex and include corrections for non-LTE effects. The errors of $A$(Li) were calculated by propagating the uncertainties in  
\teff\ and \WLi.
These final abundances are listed in Table~\ref{Tab:lithium}.

On the basis of the \teff\ (Sect.~\ref{Subsec:APs}) and \WLi\ measurements, we estimated the cluster age by using the \eagles\ code \citep{Jeffries2023}. This code fits Li-depletion 
isochrones to the \teff\ (in the range 3000--6500\,K) and \WLi\ values of a coeval stellar group. We note that \eagles\ performs optimally for clusters exhibiting a wide \teff\ 
distribution among their members, reaching out the coldest members, where the Lithium depletion is more pronounced. Unlike nearby clusters, like the Pleiades \citep[e.g.,][]{Frasca2025}, 
the useful \lamost\ MRS spectra of the NGC\,1647 members are limited to stars as cool as the Sun, with only a handful of stars extending down to $\approx$\,5300\,K. However, this 
temperature range is sufficient to obtain a reliable estimate of the age of NGC\,1647. For this analysis, we selected the stars cooler than 6500\,K, including also the few \WLi\ upper limits (a total of 73 objects). The results of applying EAGLES to our data are shown 
in Fig.~\ref{fig:EAGLES}. We find an age of 203\,$\pm$\,27\,Myr.

\begin{figure}
\begin{center}
\includegraphics[width=8.5cm,viewport= 10 10 420 320]{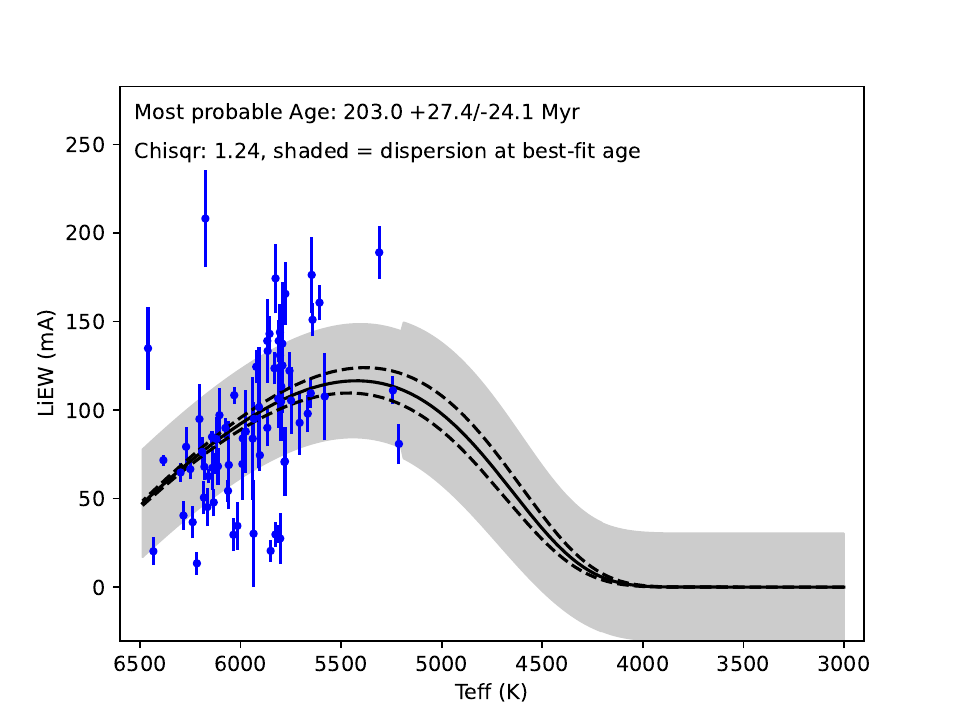}
\caption{Fit to the lithium depletion pattern of the \WLi\ measured in this work obtained with the \eagles\ code \citep{Jeffries2023}.
}
\vspace{-0.7cm}
\label{fig:EAGLES}
\end{center}
\end{figure}

\subsection{Gyrochronology}\label{Subsec:gyro}
Another helpful diagnostic tool for cluster age estimation is the distribution of rotation periods when plotted against stellar mass or a color index 
\citep[e.g.,][and references therein]{Barnes2003}.
To this end,  as mentioned in Sect.\,\ref{Subsec:Obs_photo}, we used the high-precision, high-cadence photometric \textit{TESS} data to measure the rotation period for the largest 
possible number of cluster members. In total, we analyzed the light curves of 637 members of NGC\,1647 by means of a Lomb--Scargle periodogram to search for the highest frequencies 
in their power spectra. We focused on the low-frequency domain of the power spectrum ($f\leq 10\,{\rm d}^{-1}$), which contains the power excesses typically associated with rotational 
modulations caused by starspots on the stellar photospheres.
For each star, we visually inspected the periodogram and the phase-folded light curve. By combining the periodic signals with the stars' effective temperatures, surface gravities, 
and their locations on the CMD, we identified a sample of 160 stars whose variability is confidently attributed to stellar rotation. These stars occupy the lower main sequence and exhibit 
a single, well-defined peak in the periodogram, characteristic of spot-induced rotational modulation.  The uncertainties in \prot\ were conservatively estimated from the
half-width at half-maximum of the corresponding periodogram peak and are reported alongside the periods in Table~\ref{Tab:Prot}.

We compared the rotation periods derived in this work with those reported by \citet{Long2023}. We identified 15 stars in common between the two samples, and their rotation periods show 
excellent agreement, as can be seen in Table~\ref{Tab:Comp_Prot}. Furthermore, when comparing our results with those from \citet{Breton2025}, we found two overlapping sources whose 
rotation periods are also fully consistent.

We applied extinction corrections to obtain homogeneous intrinsic colors. Since our goal is to compare NGC\,1647 with other benchmark OCs, whose members have \gaia\ magnitudes and 
colors, we did not adopt the extinction $A_V$ values derived from our SED fitting. Instead, we used the machine-learning regression extinction map from \citet{Bai2020}, which provides 
uniform estimates of \gaia\ color excess $E(G_{\rm BP}-G_{\rm RP})$ across the sky. The intrinsic color for each star was therefore computed as
\[
(G_{\rm BP}-G_{\rm RP})_{0} = (G_{\rm BP}-G_{\rm RP}) - E(G_{\rm BP}-G_{\rm RP}).
\]
After obtaining the de-reddened colors, we then constructed the color-period diagram (CPD) for NGC\,1647 using the de-reddened $(G_{\rm BP}-G_{\rm RP})_{0}$ colors and the derived 
rotation periods. The results are shown as black crosses in Fig.~\ref{fig:gyro}. For comparison, we overplotted the rotation sequences of two benchmark clusters: the 125~Myr Pleiades 
(orange points, \citealt{Rebull2016,Frasca2025}) and the 300~Myr NGC~3532 (blue points, \citealt{Barnes2003,Fritzewski2021}). The rotation-period distribution of NGC~1647 falls between 
these two reference clusters. In particular, its upper {\it I} sequence (slow rotators) overlaps that of the Pleiades and the low-color portion of that of NGC~3532, lying in between 
the two ones. Its lower {\it C} sequence (fast rotators) is more similar to that of the Pleiades, because fast rotators are observed only for low-mass stars, 
$(G_{\rm BP}-G_{\rm RP})_{0}>1.1$, in 300-Myr-old clusters like NGC\,3532, and the {\it C} sequence totally disappears at older ages. 
This comparison indicates that NGC~1647 has an intermediate level of rotational evolution. Therefore, based on gyrochronology, we infer that the age of NGC~1647 lies between 
approximately 125~Myr and 300~Myr, in good agreement with the constraints from CMD isochrone fitting and photospheric lithium abundances. 

\begin{figure}
\begin{center}
\includegraphics[width=9cm]{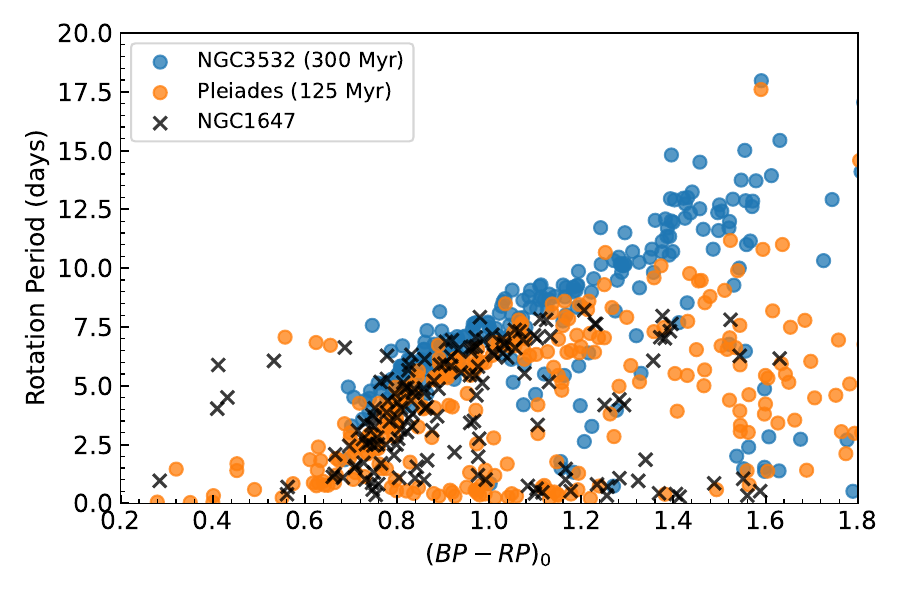}    
\vspace{-0.7cm}
\caption{Rotation periods versus the dereddened color index $(G_{\rm BP}-G_{\rm RP})_0$ for the members of NGC\,1647 (black crosses). The periods of Pleiades ($\tau\simeq125$\,Myr) 
members based on \ktwo\ \citep{Rebull2016} and \tess\ \citep{Frasca2025} are overplotted with orange dots, while those of NGC~3532 ($\tau\simeq300$\,Myr) members \citep{Fritzewski2021} 
are also shown by blue dots for comparison.
}
\vspace{-0.7cm}
\label{fig:gyro}
\end{center}
\end{figure}

\section{Discussion}
The phenomenon of eMSTO was first detected in massive Magellanic Cloud clusters and initially interpreted as a manifestation of prolonged star formation lasting several hundred 
million years \citep[e.g.,][]{Milone2009}. 
More recently, eMSTOs and broadened main sequences have been identified within Galactic OCs \citep[e.g.,][and references therein]{Cordoni2018,Cordoni2024}. 
However, no direct evidence for such long-lived star formation has been found \citep[e.g.,][]{Cabrera-Ziri2016}. Moreover, many low-mass OCs lack the escape velocity required 
to retain the gas necessary for secondary star-forming events \citep{Bastian2018}.

The most widely accepted explanation currently attributes the eMSTO to a spread in stellar rotation rates among stars of the same age \citep{Bastian_de_Mink2009,Lim2019}.
Rapid rotation causes structural changes (centrifugal support) and internal mixing, making these stars appear cooler (redder) and extending their main-sequence lifetimes. 
Furthermore, for very fast rotators, the inclination of the rotation axis plays a significant role: due to gravity darkening, the stellar poles appear hotter and brighter than 
the equatorial regions, making the observed color and magnitude dependent on the viewing angle. 
In specific cases, such as the cluster Stock\,2, the observed eMSTO is primarily caused by non-uniform interstellar extinction (differential reddening) rather than intrinsic 
stellar physics \citep{Alonso2021}. However, this does not appear to be a universal cause, as many clusters with eMSTOs show no significant reddening variations \citep{Cordoni2018}.

To investigate the primary mechanism driving the eMSTO and the broadened MS in NGC\,1647, we examined potential correlations between a star's position on the \gaia\ CMD and 
parameters such as stellar rotation and extinction. Operationally, we empirically defined a lower envelope for the MS belt (see Fig.~\ref{Fig:envelope}) using a fourth-order 
polynomial and calculated the color offset of each star relative to this locus.
\begin{figure}   
\centering            
\includegraphics[width=8.7cm,viewport= 10 0 480 360]{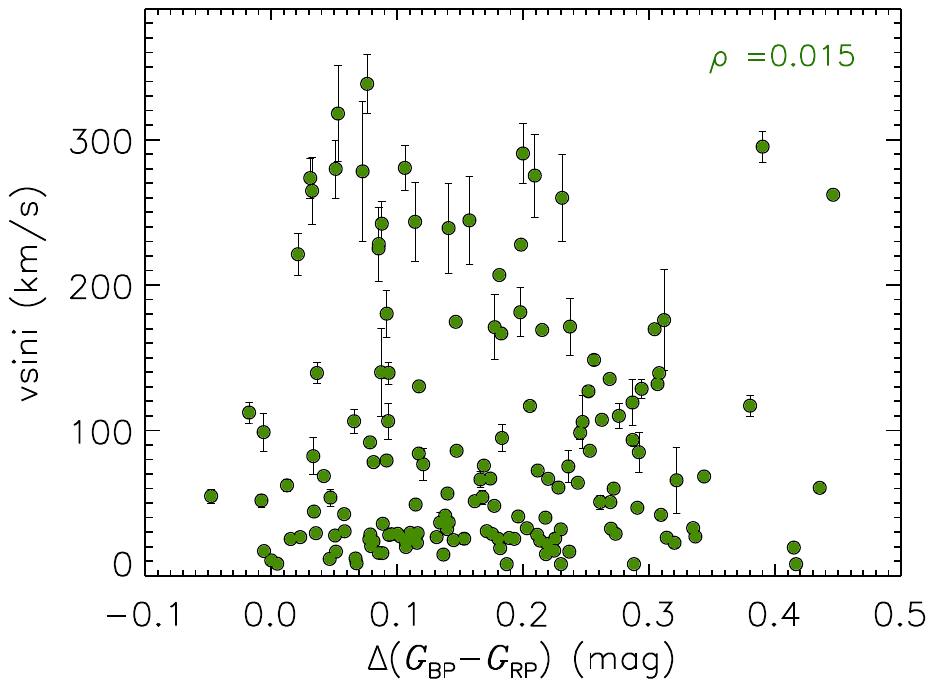}      
\caption{Projected rotational velocity (\vsini) versus the \gaia\ color shift $\Delta(G_{\rm BP}-G_{\rm RP})$. The color shift is measured relative to the lower envelope of the 
MS strip (Fig~\ref{Fig:envelope}). The Pearson's  correlation coefficients ($\rho$) is marked in the upper right corner of the panel.
}   
\label{Fig:excess_vsini} 
\end{figure}
In Fig.\,\ref{Fig:excess_vsini} we plot \vsini\ values against the observed color offset. As evidenced by the figure, and further supported by a near-zero Pearson correlation 
coefficient ($\rho=0.015$), stellar rotation does not appear to be the primary mechanism driving the MS broadening.
\begin{figure}   
\centering            
\includegraphics[width=8.7cm,viewport= 10 0 480 360]{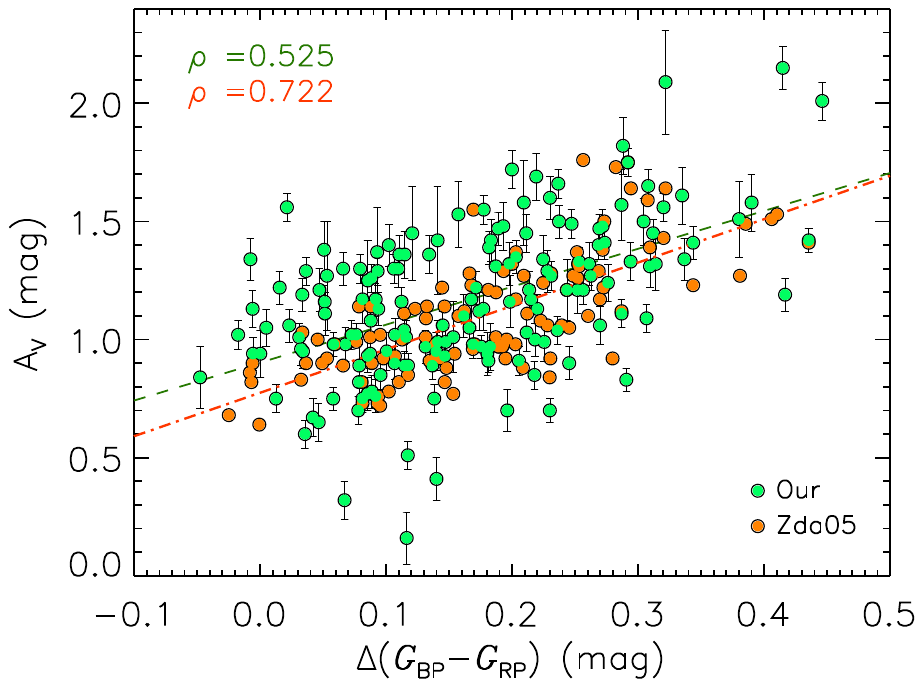}      
\caption{Extinction ($A_V$) versus the \gaia\ color shift $\Delta(G_{\rm BP}-G_{\rm RP})$. 
Data derived in the present work and by \citet{Zdanavicius2005} are distinguished by different colors, as indicated in the legend. The linear best fits to our data and 
to the \citet{Zdanavicius2005} data are shown by a green dashed line and an orange dot-dashed line, respectively. The Pearson's  correlation coefficients ($\rho$)
are also marked in the upper left corner.
}   
\label{Fig:excess_Av} 
\end{figure}
In contrast, Fig.\,\ref{Fig:excess_Av} illustrates that interstellar extinction exhibits a strong correlation with the Gaia color shift. We find Pearson correlation coefficients 
of 0.520 using our $A_V$ estimates and 0.722 using the $A_V$ values from \citet{Zdanavicius2005}. This provides compelling evidence that differential extinction accounts for most, 
if not all, of the observed MS spread.

\section{Summary}
\label{Sec:Concl}

We have presented the results of the analysis of 347 \lamost\ medium-resolution spectra of  stars that are members or candidates to the NGC\,1647 open cluster. 
This represents by far the largest sample studied to date. We were able to determine, with the analysis code \rotfit, the atmospheric parameters (\teff, \logg, and \feh), the 
radial velocity (\vrad), and the projected rotational velocity (\vsini) for most of these spectra corresponding to 158 unique stars. For four additional sources that we discovered as 
new double-lined spectroscopic binaries (SB2s), we report the radial velocities of the two components but did not determine their parameters.

We found a radial velocity distribution with a nearly symmetric peak centered at $-5.32$\,\kms\ with a dispersion $\sigma$\,=\,1.57\,\kms. Some discrepant \vrad\ values 
(by more than 5$\sigma$ from the peak center) are associated to rapidly rotating stars or to objects with multi-epoch spectra exhibiting \vrad\ variations (possible SB1 systems). 
However, some stars with a lower membership probability are found among them.
From the analysis of the cooler sources in our sample (\teff\,$\leq 7000$\,K) we derived an average metallicity \feh\,$=-0.08\pm0.09$\,dex, which is in good agreement with the 
few literature determinations that are essentially based on the two only giant cluster members.

Thanks to the parameters derived with \rotfit\ and to the spectral energy distributions from publicly available optical and NIR photometry, we could measure the stellar radius, 
luminosity, and interstellar extinction ($A_V$) for these sources. We found an excellent agreement between our extinction determinations and those from the literature 
\citep{Zdanavicius2005} with a wide $A_V$ distribution suggesting a differential reddening in the cluster field. We found a high correlation between $A_V$ and the red-ward shift 
of the \gaia\ color $(G_{\rm BP}-G_{\rm RP})$ with respect to the lower envelope of the main sequence (MS) in the color-magnitude diagram (CMD), while no correlation is found with 
the \vsini. This suggests the differential extinction as the mainly responsible for the broad MS.

From the lithium equivalent widths of the cluster members with \teff\ in the range 5300--6500\,K, we derived an age of  203\,$\pm$\,27\,Myr, by fitting empirical lithium 
isochrones thanks to the \eagles\ code \citep{Jeffries2023}. This age determination is in close agreement with that inferred from the CMD and Hertzsprung-Russell diagram 
(in the range 150--200\,Myr) and 
the one we got from the gyrochronology. In particular, for the last age diagnostics, we retrieved \tess\ light curves for 637 cluster members, which were analyzed with Lomb-Scargle 
periodogram technique. We found rotation periods for 160 stars and built the color-period diagram, which was compared to two reference clusters, i.e., the Pleiades (125\,Myr) and 
NGC~3532 (300\,Myr), placing NGC\,1647 in between them.  This is the first study of NGC~1647 to combine detailed spectroscopy for over a hundred members with high-precision photometry, 
establishing a crucial anchor for characterizing the eMSTO phenomenon, mapping differential extinction, and constraining the evolutionary properties of open clusters at this age.

\section{Data availability}
\label{Sec:Availability}

Tables~\ref{Tab:APs}, \ref{Tab:Halpha}, \ref{Tab:lithium}, \ref{Tab:SED}, and \ref{Tab:Prot} are only available at the CDS via anonymous ftp to {\tt cdsarc.u-strasbg.fr (130.79.128.5)} or via 
{\tt http://cdsarc.u-strasbg.fr/viz-bin/qcat?J/A+A/?/?}.

\begin{acknowledgements}
We are grateful to the anonymous referee for a careful reading of the manuscript and very useful suggestions.
Guoshoujing Telescope (the Large Sky Area Multi-Object Fibre Spectroscopic Telescope \lamost) is a National Major Scientific Project built by the Chinese Academy of Sciences. 
Funding for the project has been provided by the National Development and Reform Commission. \lamost\ is operated and managed by the National Astronomical Observatories, 
Chinese Academy of Sciences.
AF, JAS, and GC acknowledge funding from the Large Grant INAF-2024 ``Spectral Key features of Young stellar objects: Wind-Accretion LinKs Explored in the infraRed (SKYWALKER)''.
AB acknowledges funding from the INAF MiniGrant 2022 ``High resolution spectroscopy of open clusters". 
MQ, JNF, and JZ acknowledge the support of National Natural Science Foundation of China (NSFC) through the Grants 12427804, and the science research grants from the China Manned 
Space Project.
This research made use of SIMBAD and VIZIER databases, operated at the CDS, Strasbourg, France.
\end{acknowledgements}

\bibliographystyle{aa}
\bibliography{NGC1647_LAMOST.bib}

\onecolumn

\newpage

\begin{appendix}
\section{Resolution of \lamost\ MRS}
\label{Appendix:Resolution}

To measure the spectral resolution of \lamost\ MRS, we utilized the spectra from the wavelength calibration lamps (Th-Ar) and measured the full width at half maximum ($W_{\lambda}$) of 
emission lines uncontaminated by blending with neighboring lines.
We employed Gaussian fits, which accurately reproduce the instrumental profile at that position on the focal plane.
The resolving power at the wavelength of each line and for every fiber used is given by the $R_{\lambda}=\lambda/W_{\lambda}$ and is shown in Fig.~\ref{Fig:Resolution} for one 
example spectrum in the blue (left panels) and red arm (right panels).

For all blue-arm spectra, the resolution is observed to slightly improve from the bluest to the reddest edge. However, due to this low degree of non-uniformity, we chose to consider 
the average resolution of the spectrum to avoid introducing unnecessary complications into the analysis. As shown in the lower panels of Fig.~\ref{Fig:Resolution}, the average resolution 
does not change significantly along the orthogonal dimension (i.e., with the fiber number). Therefore, we calculated the inverse variance weighted mean of the $R_{\lambda}$ values across 
all fibers and adopted the weighted standard deviation as the error parameter.
For both arms we found values larger than the nominal $R_{\lambda}=7500$: specifically, $R_{\rm blue}=8200\pm200$ and $R_{\rm red}=8400\pm300$. 

\begin{figure*}[!h]  
\begin{center}
\includegraphics[width=8.6cm,viewport= 90 350 540 730]{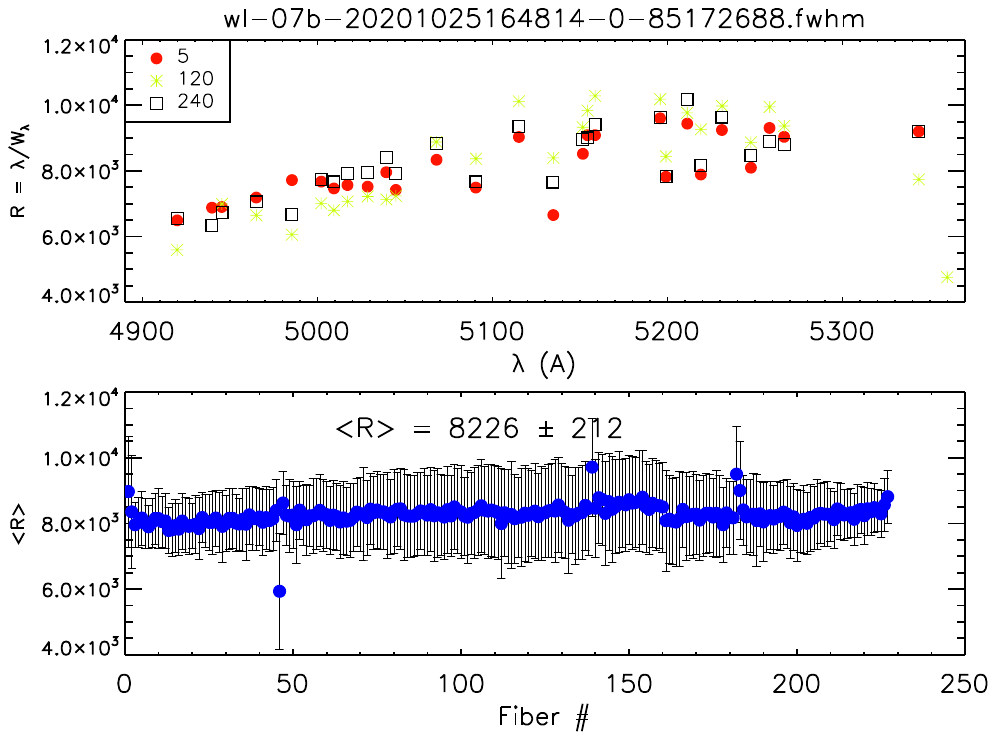}  
\includegraphics[width=8.6cm,viewport= 70 350 520 730]{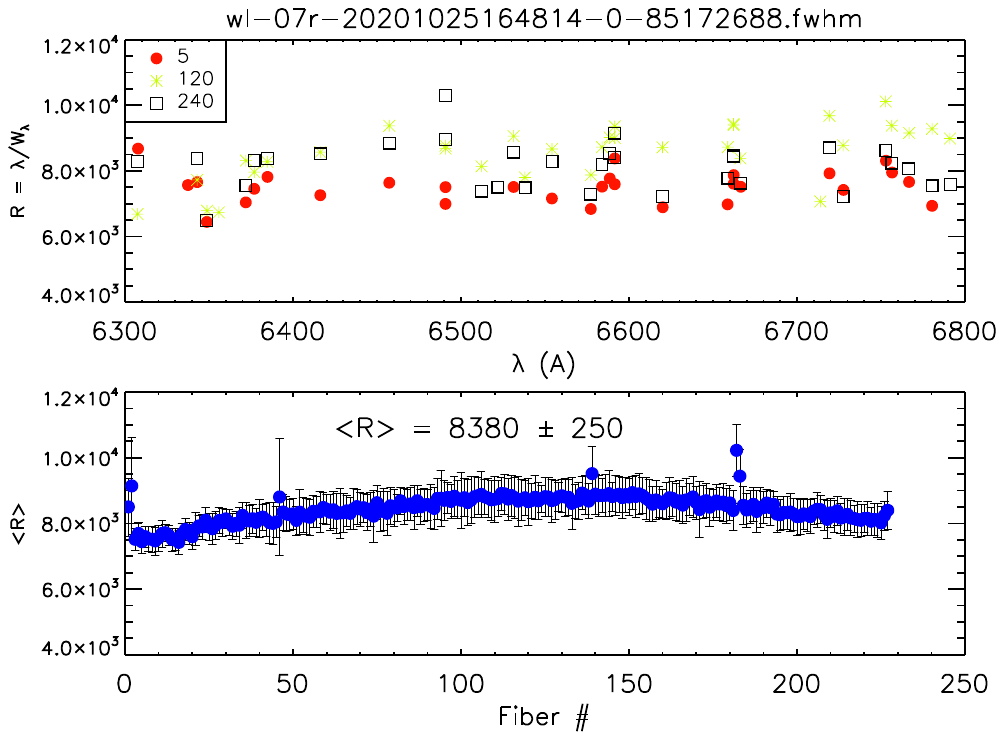}
\caption{Resolving power $R_{\lambda}=\lambda/W_{\lambda}$ of \lamost\ MRS in the blue (left panels) and red arm (right panels) as measured in an example spectrum. In the upper boxes, 
$R_{\lambda}$, for each setup, is plotted against wavelength for three fibers near the top, center, and bottom of the frame.  The lower panels display the wavelength-averaged value 
of $R_{\lambda}$ as a function of the fiber number. The mean $R$ and its uncertainty is also reported in each of the lower boxes. }
\label{Fig:Resolution}
\end{center}
\end{figure*}

\section{Additional tables and figures}
\label{Appendix:TabFig}
In this Appendix, we provide supplementary figures and tables referenced in the main text to enhance the readability of the manuscript.

\begin{figure*}[htb]
\includegraphics[width=10cm,viewport= 0 0 500 390]{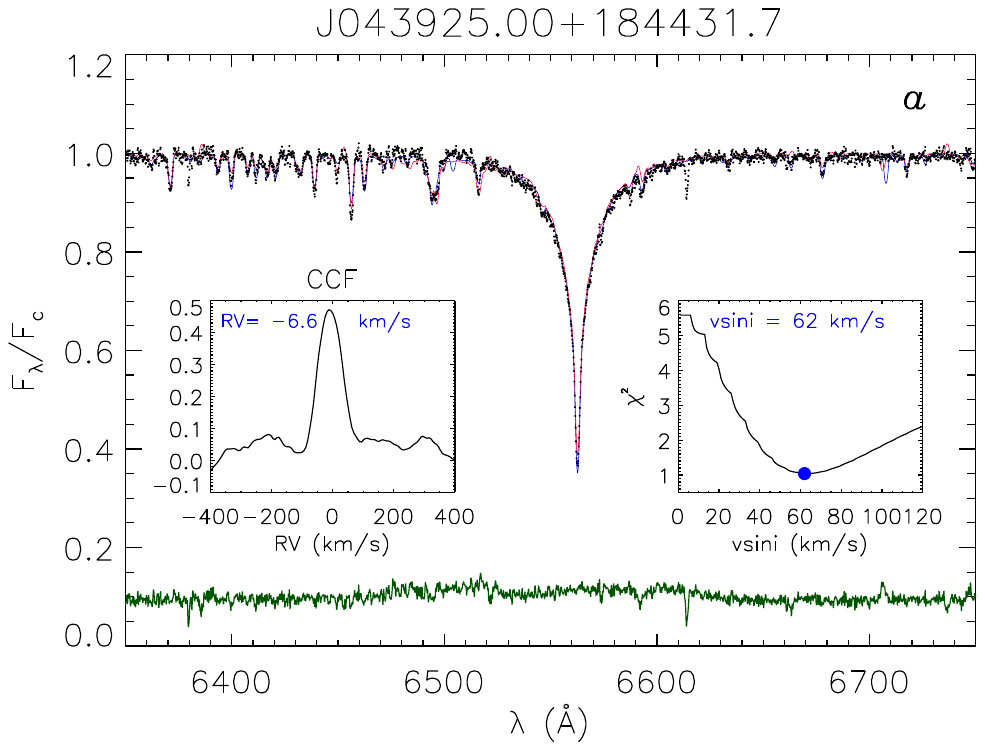}
\includegraphics[width=9.5cm,viewport= 0 0 500 390]{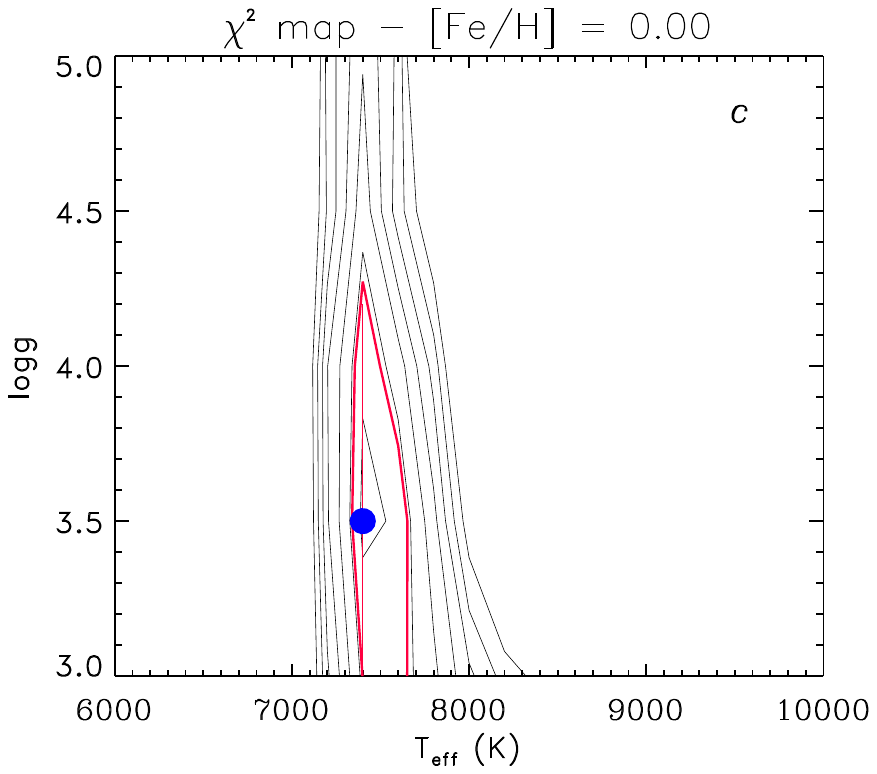}
\includegraphics[width=10cm,viewport= 0 0 500 360]{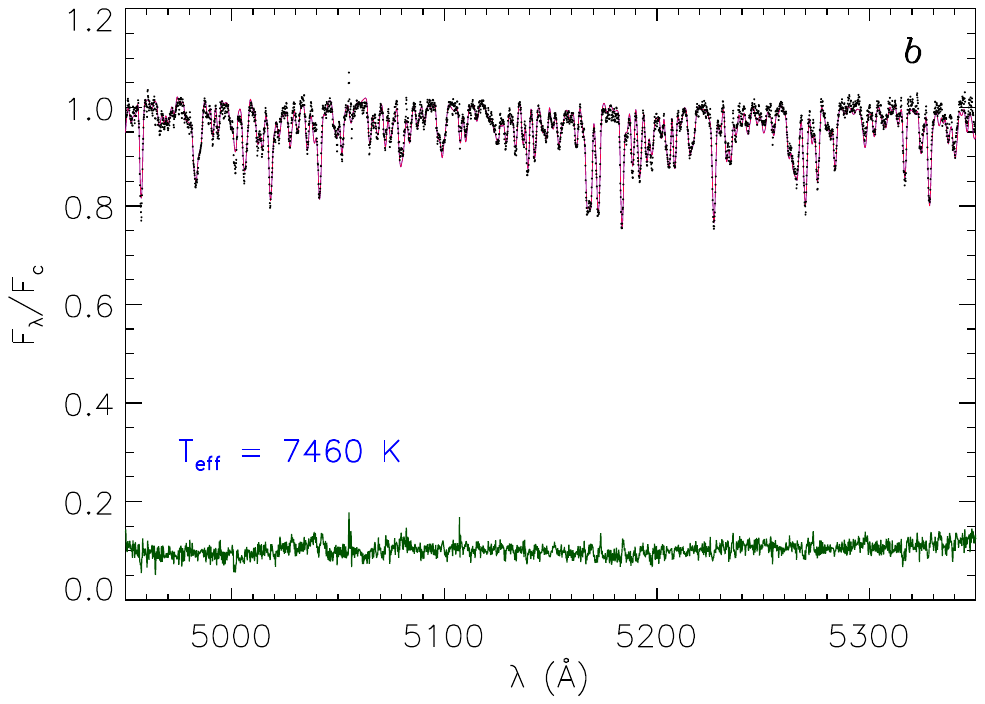}
\includegraphics[width=9.5cm,viewport= 0 0 500 360]{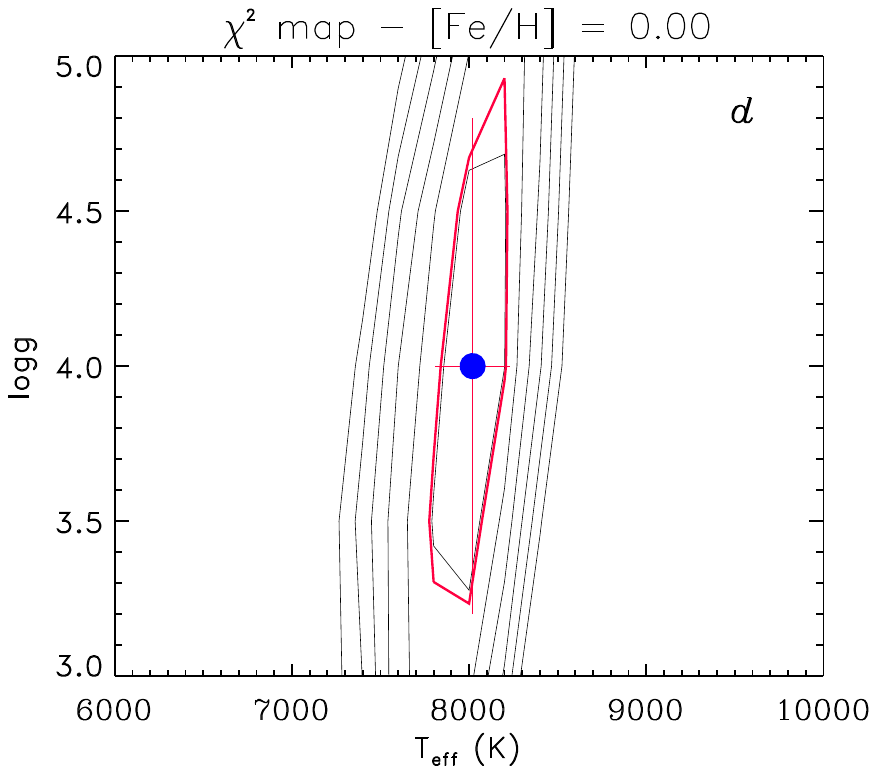}
\caption{Example of the application of the code ROTFIT to the continuum-normalized spectrum (black dots) of J043925.00+184431.7 (= $Gaia$\,DR3 3409724027479793664). 
The red-arm and blue-arm spectrum is displayed in the left-hand upper (a) and lower (b) panels, respectively. In each panel the ELODIE template spectrum broadened at the \vsini\ of 
the target is overlaid with a full red line, while the best BT-Settl template is reproduced by a blue line. The difference observed--template is shown by a dark-green line shifted 
upwards by 0.1 for the sake of clarity. The insets in panel a show the cross-correlation function (CCF) and the $\chi^2$ versus \vsini. The right-hand boxes (c and d) display the 
$\chi^2$ maps in the \teff\--\logg\ plane for the red-arm and blue-arm spectrum, respectively. The 1$\sigma$ contour is displayed by a red line, while the best-fitting parameters 
are marked with a blue dot in each panel.}
\label{Fig:ROTFIT_output}
\end{figure*}

\begin{table*}[htb]
\caption{Radial velocities of the four new SB2 systems.}
\begin{tabular}{cccrcrc}
\hline
\hline
\noalign{\smallskip}
Designation & \gaia-DR3 & HJD            &  \vrad$_{\rm 1}$   &  $\sigma_{V_r1}$ &  \vrad$_{\rm 2}$   &  $\sigma_{V_r2}$  \\
            &           & (2\,400\,000+) &  \multicolumn{2}{c}{(\kms)} & \multicolumn{2}{c}{(\kms)} \\
 \hline
\noalign{\smallskip}
 J043753.27+194239.1 & 3410615563906786048 & 58503.03992 &   4.48 &  0.46 & -49.59 & 0.95 \\
 J044405.99+184220.3 & 3409504232528828544 & 60310.11686 &   6.49 &  7.13 & -21.25 & 3.88 \\
 J044405.99+184220.3 & 3409504232528828544 & 60335.03055 &   8.28 &  0.82 & -52.35 & 1.24 \\
 J044647.31+185836.7 & 3409895452510339328 & 59599.06921 &  33.56 &  1.21 & -53.74 & 1.20 \\
 J045157.44+202337.3 & 3411603165867005568 & 59863.36632 &   9.55 &  0.43 & 103.49 & 1.02 \\
\noalign{\smallskip}
\hline \\
\end{tabular}
\label{Tab:SB2}
\end{table*}

\twocolumn

\begin{figure}[htb]
\includegraphics[width=9.5cm,viewport= 0 0 520 530]{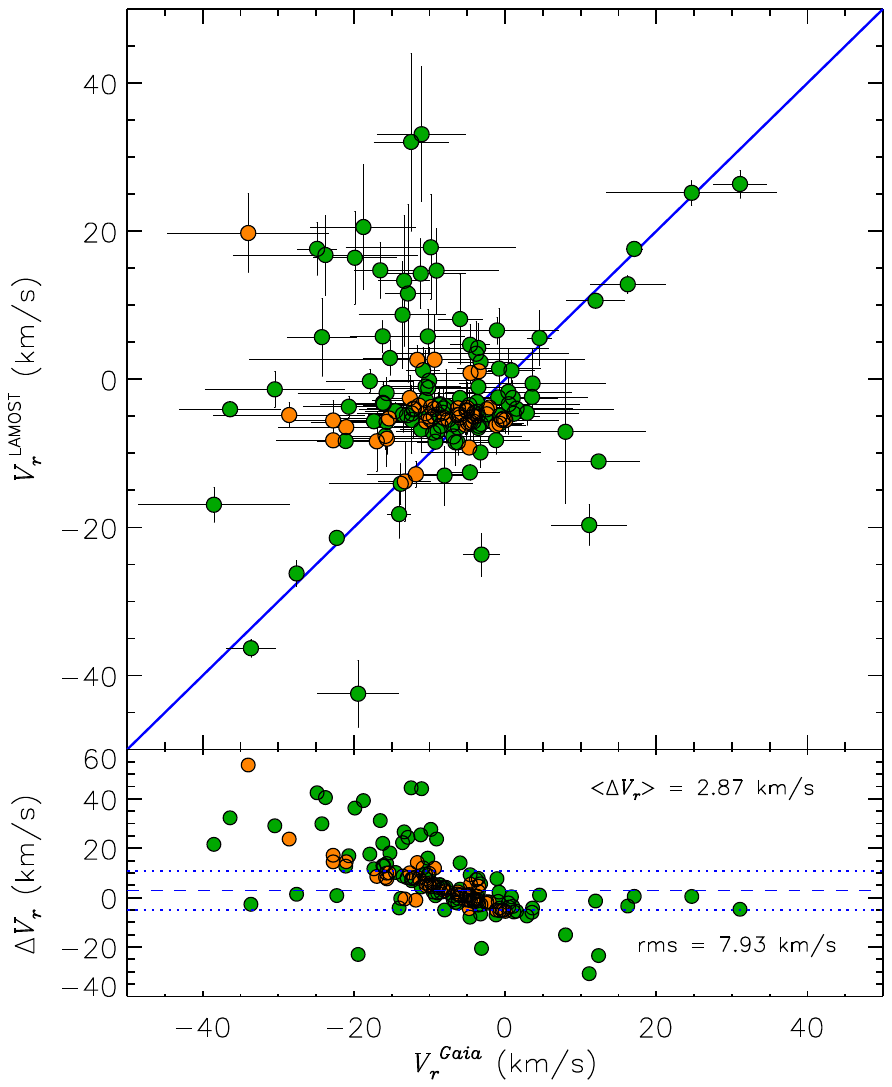}
\caption{Comparison between the average radial velocities measured in this paper with those reported in the \gaia\ DR3 catalog. Orange dots denote stars with genuine radial velocity 
variations labeled as `RVvar' in Table~\ref{Tab:APs}. The one-to-one relation is shown by the solid blue line. 
The differences $\Delta$\vrad\,=\,$V_r^{\rm LAMOST}-V_r^{Gaia}$ are displayed in the bottom panel along with their average, $<\Delta$\vrad$>$, and standard deviation, rms.
}
\label{Fig:Comp_RV_Gaia}
\end{figure}

\begin{table*}[htb]
\caption{Results of the SED fitting.}
\begin{tabular}{lclrcc} 
\hline\hline
\noalign{\smallskip}
Designation & \gaia-DR3 & ~~~\teff\ & $R$~~~~~       & $L$	    & $A_V$  \\
            &           & ~~(K)     & (R$_{\sun}$)~~ & (L$_{\sun}$) & (mag)  \\
\hline
\noalign{\smallskip}
HD\,30233$^a$  & 3410117313342257664 & 4696\,$\pm$\,11  & 43.67\,$\pm$\,0.86 & 838.0\,$\pm$\,23.6 &  1.60\,$\pm$\,0.09 \\  
HD\,30179$^a$  & 3409869064229856128 & 4914\,$\pm$\,9  & 52.54\,$\pm$\,1.16 & 1444.8\,$\pm$\,36.0 &  1.42\,$\pm$\,0.08 \\  
\noalign{\smallskip}
HD\,30123$^b$   & 3410108311089573376 & 12500\,$\pm$\,200 & 6.049\,$\pm$\,0.102 &  804.48\,$\pm$\,51.97 & 1.48\,$\pm$\,0.07 \\
HD\,284839$^b$  & 3409918164297368704 & 12000\,$\pm$\,250  &  3.032\,$\pm$\,0.039 & 171.67\,$\pm$\,14.02 & 0.94\,$\pm$\,0.06  \\
HD\,284841$^b$  & 3410107142858469248 & 11750\,$\pm$\,250 & 4.132\,$\pm$\,0.061 & 293.08\,$\pm$\,24.92 & 1.27\,$\pm$\,0.06 \\
\noalign{\smallskip}
\scriptsize{J043417.1+211726.5} &   144755932773284224  &  5816\,$\pm$\,45  &  0.967\,$\pm$\,0.028 & 0.966\,$\pm$\,0.044 &  1.31\,$\pm$\,0.09 \\ 
\scriptsize{J043735.7+192538.6} &  3410572717313072768  &  5799\,$\pm$\,42  &  1.056\,$\pm$\,0.015 &   1.136\,$\pm$\,0.041 &  0.75\,$\pm$\,0.05 \\ 
\scriptsize{J043753.3+194239.1} &  3410615563906786048  &  5804\,$\pm$\,30  &  0.998\,$\pm$\,0.015 &   1.015\,$\pm$\,0.029 &  0.98\,$\pm$\,0.07 \\ 
\scriptsize{J043831.0+191554.1} &  3410519734596650624  &  6705\,$\pm$\,41  &  1.353\,$\pm$\,0.021 &   3.342\,$\pm$\,0.104 &  0.95\,$\pm$\,0.07 \\ 
\scriptsize{J043925.0+184431.7} &  3409724027479793664  &  7462\,$\pm$\,66  &  1.548\,$\pm$\,0.022 &   6.684\,$\pm$\,0.101 &  0.75\,$\pm$\,0.06 \\ 
\scriptsize{J044022.8+192641.0} &  3410540487878678144  &  5865\,$\pm$\,28  &  1.518\,$\pm$\,0.025 &   2.464\,$\pm$\,0.074 &  0.70\,$\pm$\,0.05 \\ 
\scriptsize{J044030.8+203400.9} &  3411045644751959808  &  5922\,$\pm$\,106  &  1.074\,$\pm$\,0.013 &  1.276\,$\pm$\,0.067 &  1.10\,$\pm$\,0.09 \\ 
\hline
\noalign{\smallskip}
\end{tabular}
\\{\bf Notes.} The full table is available at the CDS.\\ $^{(a)}$ \teff\ from \citet{Carrera2022}.  $^{(b)}$ UVES archival spectra.
\label{Tab:SED}
\end{table*}

\begin{table}[ht]
\caption{Rotation periods derived in the present work.}
\begin{tabular}{crrr}
\hline
\hline
\noalign{\smallskip}
\gaia-DR3           & TIC         & \prot & error  \\
                    &             &  (d)  &  (d)   \\ 
\hline
\noalign{\smallskip}
3409916755548128512 & 18442570    & 6.937 & 0.718  \\
3410105734109204736 & 18442683    & 7.102 & 0.867  \\
3410105253072863104 & 18442693    & 3.323 & 0.465  \\
3410107967492188928 & 18442841    & 6.387 & 0.608  \\
3406767204490489856 & 18582731    & 4.875 & 1.164  \\
3410125284800199296 & 18443108    & 7.022 & 3.080  \\
3410282892919952000 & 18478360    & 6.205 & 1.508  \\
3410227779899638912 & 18478543    & 0.997 & 0.041  \\
3410118412852640128 & 18478687    & 3.040 & 0.382  \\
3410113426393937664 & 18478784    & 2.746 & 0.302  \\
3410107658254539264 & 18478797    & 2.506 & 0.302  \\
3410107314657156096 & 18478822    & 4.362 & 0.808  \\
3410107349016895616 & 18478824    & 4.884 & 1.020  \\
3409918851492142976 & 18478892    & 7.058 & 2.079  \\
3409914590884594176 & 18478990    & 4.291 & 0.723  \\
\noalign{\smallskip}
\hline \\
\end{tabular}\\
{\bf Notes.} The full table is available at the CDS.
\label{Tab:Prot}
\end{table}

\begin{table}[ht]
\caption{Rotation periods derived in the present work and in the literature.}
\begin{tabular}{crrr}
\hline
\hline
\noalign{\smallskip}
\gaia-DR3        & \prot$^{T}$  & \prot$^{L}$  & \prot$^{B}$ \\
                 &  (d)         &  (d)         &   (d)       \\
\hline
\noalign{\smallskip}
3409890710866487040 &  4.505$\pm$0.305  &  4.526$\pm$0.669  &  \dots   \\
3409912735458739328 &  5.883$\pm$0.435  &  5.925$\pm$0.632  &  \dots   \\
3410105631029990400 &  4.027$\pm$0.241  &  4.042$\pm$0.837  &  \dots   \\
3409916407654135680 &  4.669$\pm$0.272  &  4.684$\pm$0.378  &  4.753$\pm$0.760   \\
3409916755548128512 &  6.937$\pm$0.718  &  7.000$\pm$1.993  &  \dots   \\
3410085839820608384 &  3.813$\pm$0.736  &  3.800$\pm$0.609  &  \dots   \\
3410105734109204736 &  7.102$\pm$0.867  &  7.173$\pm$0.424  &  \dots   \\
3410105935971159424 &  5.524$\pm$0.453  &  5.519$\pm$0.420  &  \dots   \\
3410107967492188928 &  6.387$\pm$0.608  &  6.396$\pm$1.115  &  \dots   \\
3409892252758068864 &  1.313$\pm$0.051  &  1.320$\pm$0.149  &  \dots   \\
3409890161110686208 &  6.808$\pm$0.693  &  6.830$\pm$1.195  &  \dots   \\
3409912185702940672 &  6.459$\pm$0.628  &  6.499$\pm$0.345  &  6.516$\pm$1.077   \\
3410104394079308416 &  7.627$\pm$0.716  &  7.604$\pm$2.560  &  \dots   \\
3410111609624342144 &  1.728$\pm$0.044  &  1.727$\pm$0.180  &  \dots   \\
3410187441566870144 &  6.307$\pm$0.589  &  6.360$\pm$1.343  &  \dots   \\
\noalign{\smallskip}
\hline \\
\end{tabular}\\
{\bf Notes.}  \prot$^{T}$ = period measured in this work with \tess;  \prot$^{L}$ = period measured by \citet{Long2023} with \ktwo; \prot$^{B}$ = period measured by \citet{Breton2025} 
with \ktwo.
\label{Tab:Comp_Prot}
\end{table}

\begin{figure}   
\centering            
\includegraphics[width=\columnwidth,viewport= 60 100 570 730]{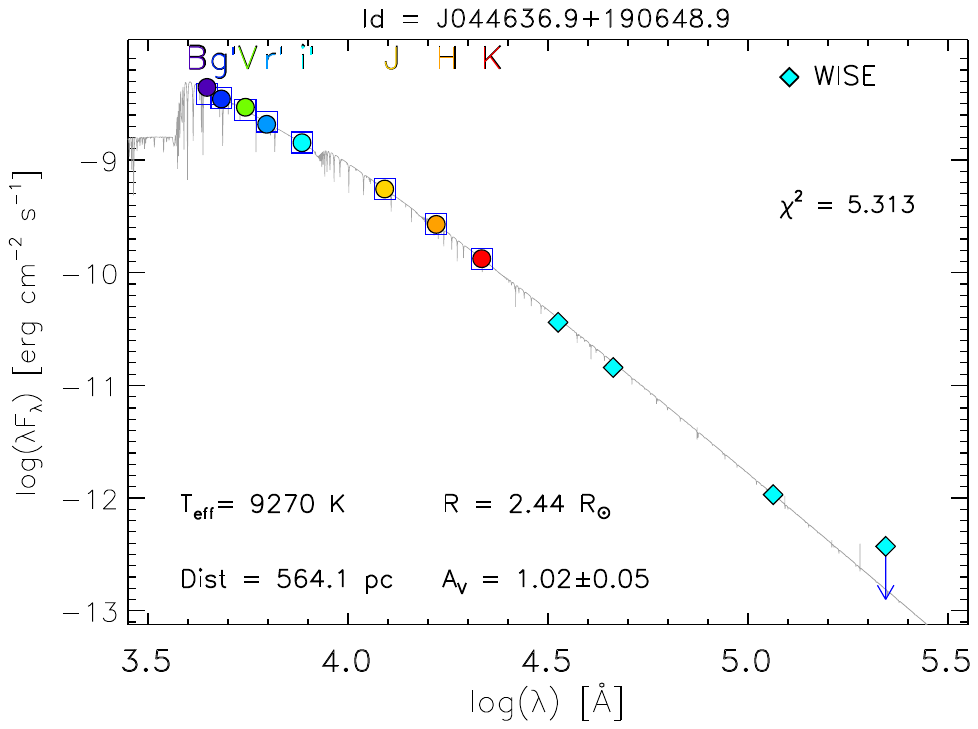}      
\includegraphics[width=\columnwidth,viewport= 60 350 570 470]{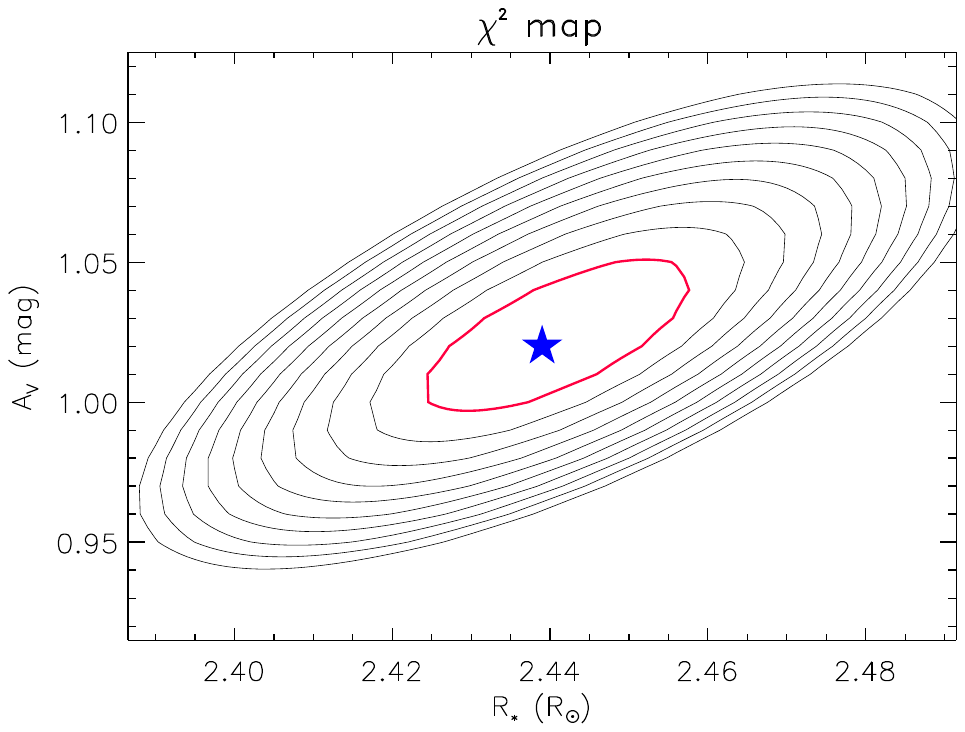} 
\caption{{\it Top:} Example of an SED fitting for the hot star J044636.9+190649 (= HD\,285997). {\it Bottom:} $\chi^2$-contour map of the fitting. The red contour corresponds to 
the 1$\sigma$ confidence level, while the best-fitting parameters (extinction and radius) are marked with a blue star symbol.  }
\label{Fig:sed1} 
\end{figure}

\begin{figure}   
\centering            
\includegraphics[width=\columnwidth,viewport= 60 100 570 730]{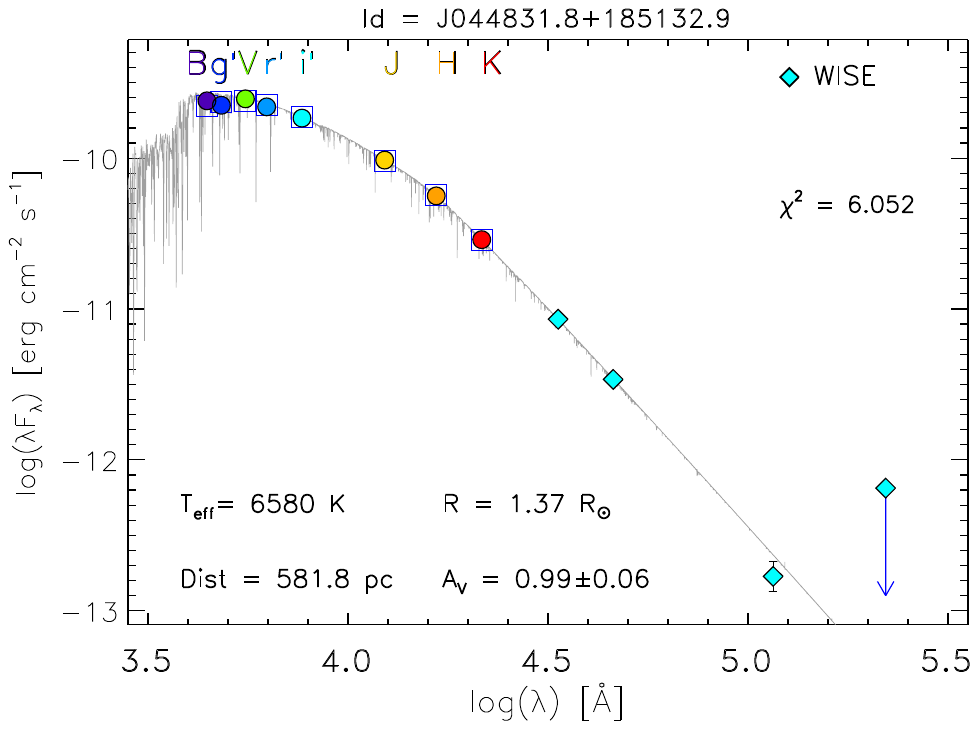}      
\includegraphics[width=\columnwidth,viewport= 60 350 570 470]{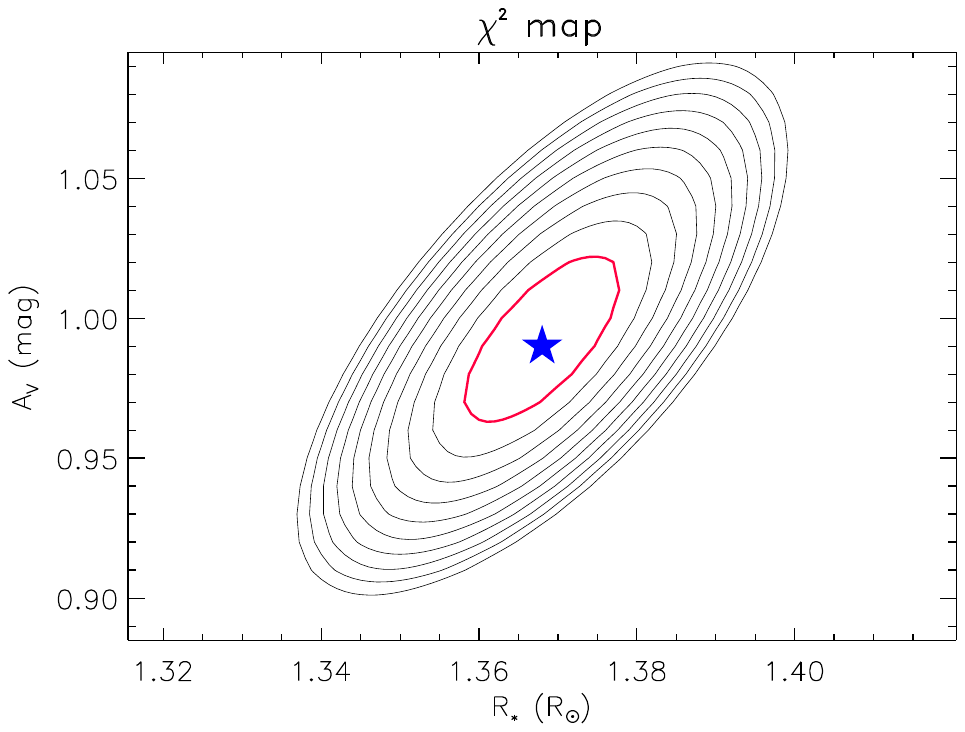}      
\caption{{\it Top:} Example of an SED fitting for the mid-F type star J044831.8+185133 (= UCAC4 545-011022). {\it Bottom:} $\chi^2$-contour map of the fitting. The red contour corresponds to the 1$\sigma$ confidence level, while the best-fitting parameters (extinction and radius) are marked with a blue star symbol.}   
\label{Fig:sed2} 
\end{figure}

\begin{figure}   
\centering            
\includegraphics[width=\columnwidth,viewport= 10 10 460 460]{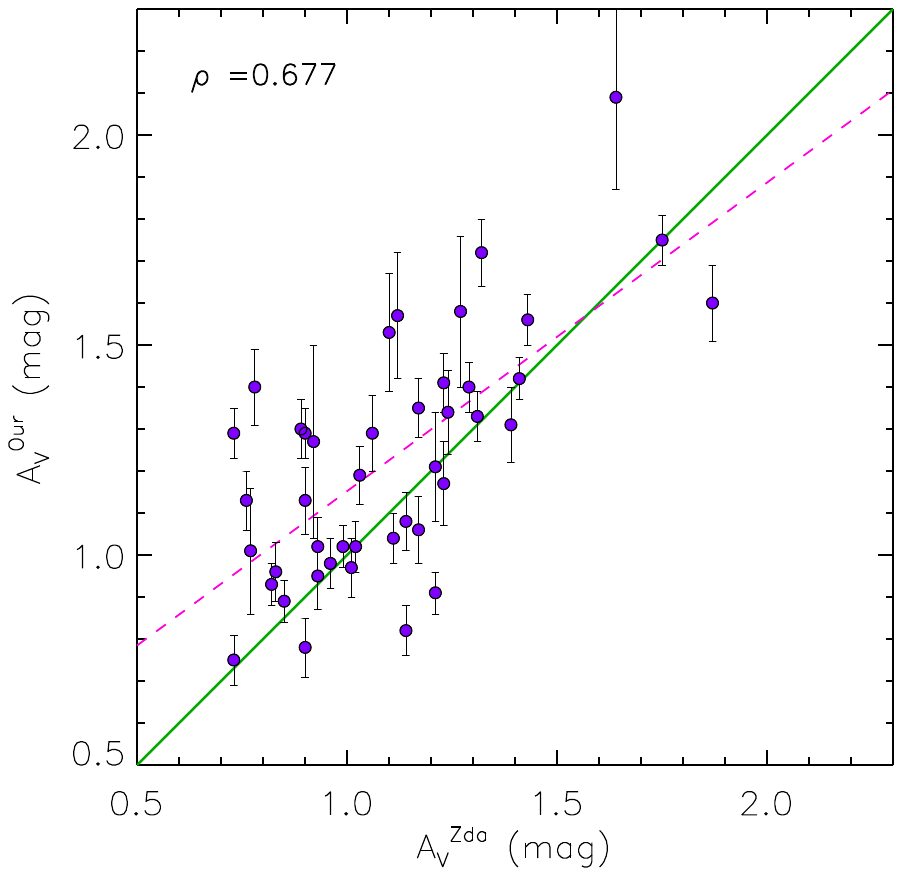}      
\caption{Comparison between the extinction values measured by us and by \citet{Zdanavicius2005}. 
The one-to-one relation is displayed with a continuous green line, while the linear best fit to the data is shown with a magenta dashed line.
}   
\label{Fig:Av_comp} 
\end{figure}

\begin{figure}   
\centering            
\includegraphics[width=\columnwidth,viewport= 20 0 370 370]{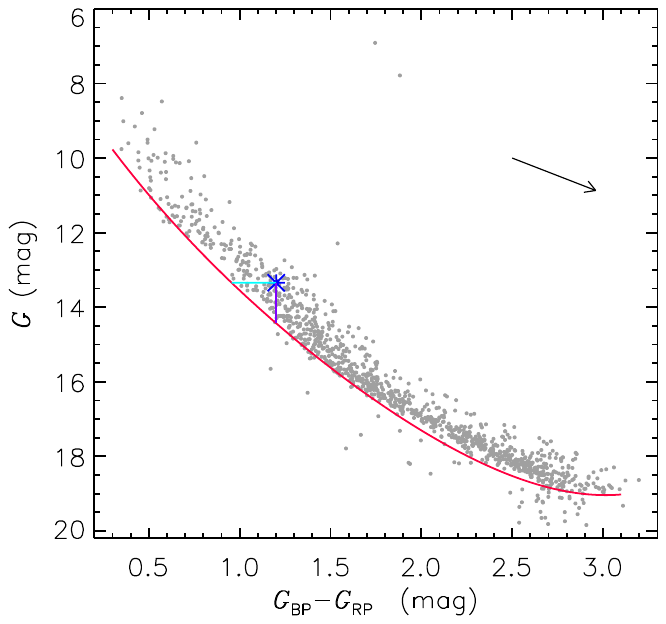}      
\caption{ \gaia\ color-magnitude diagram (CMD) of the cluster members (gray dots) showing the empirically defined lower envelope of the main sequence (red line). The color shift, 
$\Delta(G_{\rm BP}-G_{\rm RP})$, for a star (indicated by a blue asterisk) is shown by the cyan horizontal segment, while the magnitude difference ($\Delta G$) relative to this 
envelope is marked with a purple vertical segment. The black arrow indicates the direction and length of the extinction vector for a reference value of $A_V=1.1$\,mag.
}   
\label{Fig:envelope} 
\end{figure}

\end{appendix}

\end{document}